\documentclass[11pt]{article}

\usepackage[a4paper,margin=1in]{geometry}
\usepackage{amsmath,amssymb,amsfonts,amsthm}
\usepackage{graphicx}
\usepackage{hyperref}
\usepackage{bm}
\usepackage{enumerate}
\usepackage{mathrsfs}

\usepackage{tikz}
\usepackage{pgfplots}
\usetikzlibrary{arrows,shapes,positioning,calc}
\usepackage{subcaption}

\hypersetup{
  colorlinks=true,
  linkcolor=blue,
  citecolor=blue,
  urlcolor=blue
}

\newtheorem{thm}{Theorem}[section]
\newtheorem{prop}[thm]{Proposition}
\newtheorem{lemma}[thm]{Lemma}
\newtheorem{cor}[thm]{Corollary}
\newtheorem{assump}[thm]{Assumption}
\newtheorem{define}[thm]{Definition}

\numberwithin{equation}{section}

\title{
A Theoretical Framework for the Formation of Large Animal Groups:\\[0.3em]
Topological Coordination, Subgroup Merging, and Velocity Inheritance
}

\author{
Jidong Jin\thanks{Department of Computer Science and Technology,
Capital University of Economics and Business, Beijing 100070, China.}
}

\date{November 28, 2025}

\begin{document}

\maketitle

\begin{abstract}
Large animal groups---bird flocks, fish schools, insect swarms---are often assumed to form
by gradual aggregation of sparsely distributed individuals. Using a mathematically precise
framework based on time-varying directed interaction networks, we show that this widely held
view is incomplete.

The theory demonstrates that large moving groups do not arise by slow accumulation; instead,
they emerge through the rapid merging of multiple pre-existing subgroups that are simultaneously
activated under high-density conditions. The key mechanism is topological: the long-term
interaction structure of any moving group contains a single dominant strongly connected
component (SCC). This dominant SCC determines the collective velocity---both speed and
direction---of the entire group.

When two subgroups encounter one another, the trailing subgroup aligns with---and ultimately
inherits---the velocity of the dominant SCC of the leading subgroup. Repeated merging
events naturally generate large groups whose speed is predicted to be lower than the mean
speed of the original subgroups. The same dynamics explain several universal empirical
features: broad neighbour-distance distributions, directional asymmetry in neighbour selection,
and the characteristic narrow-front, wide-rear geometry of real flocks.

The framework yields testable predictions for STARFLAG-style 3D datasets, offering a unified
explanation for the formation, maintenance, and geometry of coordinated animal groups.
\end{abstract}

\tableofcontents

\bigskip

\part{Biological Narrative}

\section{Introduction}

Animals with limited sensory and cognitive abilities can nonetheless form strikingly coherent
groups. Starling flocks, fish schools, migratory bird formations, escaping insect swarms---all
show strong global organization arising from simple, local interactions.

Over the past twenty years, high-resolution tracking, especially the STARFLAG project
\cite{Ballerini2008}, has made it possible to measure these interactions directly. Several key
empirical facts now appear well-established:
\begin{itemize}
  \item Individuals do not use metric distances; they coordinate with a fixed number of
  topological neighbours.
  \item Interaction networks remain stable over long time windows, even though instantaneous
  contacts change from moment to moment.
  \item Ordered groups display strong internal geometry: narrow front, wide rear, layered density.
\end{itemize}

What remains unclear is how large coherent groups arise in nature. STARFLAG and similar
datasets capture groups after they are already coordinated; they do not capture how such groups
emerge from disordered starting conditions.

Classical models (Vicsek-type, metric zones, vision cones) can produce coordinated motion,
but they do not explain several key features:
\begin{itemize}
  \item the stability of a leading region whose influence propagates backward,
  \item the highly asymmetric geometry of real flocks,
  \item the apparent inheritance of velocity when subgroups merge.
\end{itemize}

Here we introduce a topological coordination theory that addresses precisely these gaps. The
framework begins with a simple but empirically grounded rule: individuals adjust their velocity
based on the states of their topological neighbours. Crucially, real animals do this one neighbour
at a time, but the mathematical limit of such sequential updates can be written as a continuous-time
equation.

The central theoretical result is that the long-term interaction structure of any moving group
always contains a single dominant subgroup, mathematically a strongly connected component
(SCC) that receives no persistent influence from others. This dominant SCC dictates the final
velocity of the entire group.

This leads to a simple biological interpretation:
\begin{itemize}
  \item A subgroup that is not influenced by others becomes the leader.
  \item When two subgroups encounter one another, the trailing group inevitably aligns with the
  dominant subgroup of the group in front.
  \item The trailing group's speed therefore decreases---a mechanism we call \emph{velocity
  inheritance}.
  \item Repeated encounters cause multiple subgroups to merge, producing large groups without
  requiring gradual aggregation.
\end{itemize}

This mechanism naturally predicts:
\begin{itemize}
  \item broad, unstructured distributions of neighbour distances;
  \item directionally asymmetric neighbour selection;
  \item and the characteristic narrow-front, wide-rear geometry of real flocks.
\end{itemize}

It also predicts that very large groups must move more slowly than small groups, a hypothesis
that can be directly tested using existing datasets.

Overall, this work provides a unified, data-driven framework linking topology, subgroup
structure, and collective motion, and it offers a mechanistic explanation for how large coordinated
animal groups form in the wild.

\section{Asymmetric Velocity Coordination System}

\subsection{Equations of Motion}

To describe how animals move and coordinate in real physical space, the model must respect
a basic physical constraint: an animal cannot change its velocity instantaneously. Any change of
speed or turning requires the production of a force, which in turn generates acceleration. Thus
a continuous-time description naturally matches the way animals adjust their motion in the real
world.

A second point is often overlooked. In collective behaviour studies, ``turning'' is usually treated
as a change in direction, but physically it is also a form of acceleration. Whether an individual
speeds up, slows down, or turns, it must exert force, and the resulting acceleration accumulates
over time. None of these adjustments occur in a single instant.

Let $\mathbb{A}$ denote the group of individuals, and let $a_i \in \mathbb{A}$ ($i = 1, \dots, n$)
be one such individual. The motion state of $a_i$ is represented by its velocity vector
$v_i \in \mathbb{R}^m$ ($m = 1,2,3$). When individual $a_i$ responds to neighbour $a_j$, it
generates a force that drives its own change of motion:
\[
  \bm{f}_i = \bm{f}_i(v_i, v_j).
\]
Because each animal perceives the world from its own moving coordinate frame, the observable
cue is the relative motion $v_j - v_i$. Hence the force can be written as
\[
  \bm{f}_i = \bm{f}_i(0, v_j - v_i) = \bm{f}_i(v_j - v_i).
\]
Biologically, $a_i$ attempts to reduce the difference between its own motion state and that of
$a_j$. Thus the force points in the direction from $v_i$ to $v_j$, and we may express it as
\begin{equation}\label{eq:fi_S}
  \bm{f}_i = S_i(v_j - v_i)\,
  \frac{v_j - v_i}{\|v_j - v_i\|},
\end{equation}
where $S_i(\cdot)\ge 0$ measures the strength of the stimulus from $a_j$ to $a_i$. This scalar
captures how strongly individual $a_i$ responds to a given difference in motion.

Physical laws must hold regardless of the coordinate frame. Reversing all axes changes the
sign of every vector but must not change the functional form of the dynamics. Comparing the
transformed and original expressions leads to the conclusion that $S_i$ depends only on the
magnitude of the relative motion:
\[
  S_i(v_j - v_i) = S_i(-(v_j - v_i)) = S_i(\|v_j - v_i\|).
\]
Substituting this into \eqref{eq:fi_S} gives
\begin{equation}\label{eq:fi_r}
  \bm{f}_i = S_i(\|v_j - v_i\|)\,
  \frac{v_j - v_i}{\|v_j - v_i\|}.
\end{equation}
If individual $a_i$ has mass $m_i$, then
\begin{equation}\label{eq:vi_dot_basic}
  \dot{v}_i = \frac{\bm{f}_i}{m_i}
  = \frac{1}{m_i}S_i(\|v_j - v_i\|)
  \frac{v_j - v_i}{\|v_j - v_i\|}
  = g_i(\|v_j - v_i\|)
  \frac{v_j - v_i}{\|v_j - v_i\|},
\end{equation}
where
\[
  g_i(\|v_j - v_i\|) = \frac{1}{m_i}S_i(\|v_j - v_i\|).
\]
Equation \eqref{eq:vi_dot_basic} captures the physical process underlying motion adjustment.
It is invariant under translations, rotations, and reflections, and is therefore an affine dynamical
equation: its form does not depend on the observer's coordinate system.

In \eqref{eq:vi_dot_basic}, the response function $g_i(\cdot)$ carries the index $i$, reflecting
individual heterogeneity. Animals differ in body size, muscular capacity, sensory acuity, reaction
speed, or attentional state, and therefore the same stimulus may produce different accelerations
in different individuals.

To incorporate an additional layer of biological realism, we may generalise $g_i(\cdot)$ to
$g_{ij}(\cdot)$. This allows the response of $a_i$ to depend not only on its own characteristics,
but also on which neighbour $a_j$ is producing the stimulus. Animals often react differently to
adults versus juveniles, healthy versus weakened individuals, or conspecifics versus predators. The
function $g_{ij}(\cdot)$ provides a natural way to encode these differences.

Finally, by summing over all neighbours and allowing neighbour relations to change through
time, we obtain the full equation of motion for $a_i$:
\begin{equation}\label{eq:vi_full}
  \dot{v}_i(t)
  = \sum_{a_j \in N(t,a_i)}
  g_{ij}(\|v_j(t) - v_i(t)\|)
  \frac{v_j(t) - v_i(t)}{\|v_j(t) - v_i(t)\|},
\end{equation}
where
\begin{equation}\label{eq:Ni_def}
  N(t,a_i)
\end{equation}
denotes the neighbour set of individual $a_i$ at time $t$.

Although \eqref{eq:vi_full} appears to combine influences ``simultaneously'', real animals do
not integrate multiple stimuli at a single instant. Due to clear limits in attention and information
processing, individuals typically update their motion sequentially, reacting to one neighbour at
a time. Thus the empirical correlation with average neighbour velocity reported by Ballerini
et al.\ \cite{Ballerini2008} should be interpreted as the cumulative outcome of a sequence of
neighbour-specific responses, rather than a literal instantaneous averaging.

Moreover, \eqref{eq:vi_full} applies not only to birds, fish, and insects, but also to higher
vertebrates. For instance, many mammals pursued by multiple predators can exploit brief gaps
between hunters, choosing an escape direction informed by several attackers simultaneously. This
behaviour requires integration of directional cues from multiple sources, and \eqref{eq:vi_full}
accommodates such scenarios naturally.

\subsection{Velocity Coordination System}

By simply reversing the direction of the interaction vector, we can rewrite \eqref{eq:vi_full}
in the canonical form
\begin{equation}\label{eq:velocity_coordination}
  \dot{v}_i(t)
  = \sum_{a_j \in N(t,a_i)}
  -g_{ij}(\|v_i(t) - v_j(t)\|)
  \frac{v_i(t) - v_j(t)}{\|v_i(t) - v_j(t)\|}.
\end{equation}
A many-body motion system in Euclidean space is called a \emph{first-order additive many-body
dynamical system} if each moving entity satisfies an equation of the form \eqref{eq:velocity_coordination}.

By contrast, if each entity satisfies
\begin{equation}\label{eq:second_order}
  \dot{v}_i(t)
  = \sum_{a_j \in N(t,a_i)}
  -g_{ij}(\|x_i(t) - x_j(t)\|)
  \frac{x_i(t) - x_j(t)}{\|x_i(t) - x_j(t)\|},
\end{equation}
where $x_i(t)$ and $x_j(t)$ are the positions of entities $i$ and $j$, then the system is a
\emph{second-order additive many-body dynamical system}.

The classical Newtonian $N$-body system in celestial mechanics is a standard example of a
second-order additive many-body system. In particular, if in \eqref{eq:second_order} we choose
\[
  g_{ij}(\|x_i(t) - x_j(t)\|)
  = \frac{G M_i M_j}{\|x_i(t) - x_j(t)\|^2 M_i}
  = \frac{G M_j}{\|x_i(t) - x_j(t)\|^2},
\]
where $M_i$ and $M_j$ are the masses of bodies $i$ and $j$, $G > 0$ is the gravitational constant,
and the neighbour set $N(t,a_i)$ of body $i$ contains all bodies in the system, then
\eqref{eq:second_order} reduces exactly to the classical Newtonian gravitational $N$-body model.

From classical mechanics it is known that a second-order system driven purely by conservative
interaction forces is a Hamiltonian system. Its total energy is conserved, and the resulting tra-
jectories typically exhibit oscillatory or even periodic motion rather than convergence. Without
some form of energy dissipation (for example, inelastic collisions), such a system cannot reach a
common state. In other words, a purely conservative second-order model cannot generate consensus.

This observation is important when applying second-order models, or models with second-order
components, to collective animal behaviour. If no dissipative mechanisms are included, the system
remains Hamiltonian and tends to oscillate instead of converging. This limitation does not arise
from biology but from the mathematical structure of the underlying dynamics.

At the same time, second-order control is clearly present in real animal behaviour. For instance,
a lone goose accelerates to catch up with its flock, and many animals adjust their speed in response
to others. Thus second-order effects are biologically important, but a purely conservative second-
order model is inadequate. Real animals operate under non-conservative constraints such as limited
locomotor capacity, environmental damping (e.g.\ air or water resistance), and neural delays, all
of which cap their maximum speed and break the Hamiltonian nature of the system. Any realistic
second-order control model must therefore incorporate such dissipative or limiting factors.

In this paper we focus on the first-order many-body dynamical system \eqref{eq:velocity_coordination}.
In \eqref{eq:velocity_coordination}, the field-strength function $g_{ij}$ is not restricted to any
specific analytical form; it may be any sufficiently regular function. This flexibility allows two
biologically relevant regimes:
\begin{itemize}
  \item a regime in which individual locomotion limits are effectively ignored; and
  \item a regime in which individual acceleration limits are explicitly present.
\end{itemize}
In the latter case, we may assume that
\[
  0 \le |g_{ij}(x)| \le M,
\]
where $M$ represents the maximum acceleration that an individual can generate. This bound can
also absorb the effects of environmental damping. When damping is present, the effective inter-
action strength can be written as
\[
  g_{ij}(\|v_i - v_j\|, \|v_i\|)
  = g'_{ij}(\|v_i - v_j\|) - D(\|v_i\|),
\]
where $g'_{ij}$ describes the interaction in the absence of damping, and $D(\|v_i\|)$ is a damping
term.

Since we are interested in the coordination of moving animals, and animals usually travel around
their preferred cruising speeds, variations in speed $\|v_i\|$ are often small and can be approximated
as constant. Under this approximation, the interaction strength becomes
\[
  g_{ij}(\|v_i - v_j\|)
  = g'_{ij}(\|v_i - v_j\|) - D(\|v_i\|)
  = g'_{ij}(\|v_i - v_j\|) - M,
\]
where $M$ can be treated as an effective constant damping coefficient.

\begin{define}[Velocity coordination system]\label{def:velocity_coord}
A first-order many-body dynamical system of the form
\begin{equation}\label{CS1}
  \dot{v}_i(t)
  = \sum_{a_j \in N(t,a_i)}
  -g_{ij}(\|v_i(t) - v_j(t)\|)
  \frac{v_i(t) - v_j(t)}{\|v_i(t) - v_j(t)\|}
\end{equation}
is called a \emph{velocity coordination system} if, for any $i, j$, the function $g_{ij}(\|v_i - v_j\|)$
satisfies
\begin{equation}\label{CS2}
  g_{ij}(\|v_i - v_j\|)
  =
  \begin{cases}
    0, & v_i = v_j,\\[0.2em]
    > 0, & v_i \neq v_j.
  \end{cases}
\end{equation}
The quantity $g_{ij} \ge 0$ is called the \emph{coordination-strength function} of individual
$a_j$ acting on individual $a_i$. Since it depends on the Euclidean distance between the states
of $a_j$ and $a_i$, it can also be interpreted as the field intensity exerted by $a_j$ on $a_i$.
\end{define}

The subsequent analysis is based on the following regularity assumption.

\begin{assump}\label{ass:C1}
In the velocity coordination system \eqref{CS1}, all coordination-strength functions $g_{ij}$ are
assumed to be $C^1$ (continuously differentiable), or, more generally, piecewise $C^1$ functions
whose derivatives may have discontinuities of the first kind.
\end{assump}

The physical meaning of Assumption~\ref{ass:C1} can be seen by writing
\begin{equation}\label{eq:gij}
  y = g_{ij}(x).
\end{equation}
The incremental change
\[
  \frac{\Delta y}{\Delta x}
  = \frac{g_{ij}(x+\Delta x) - g_{ij}(x)}{\Delta x}
\]
is required to have a finite limit as $\Delta x \to 0$. That is, the derivative
\[
  \frac{dy}{dx} = \lim_{\Delta x \to 0} \frac{\Delta y}{\Delta x}
\]
exists and remains finite.


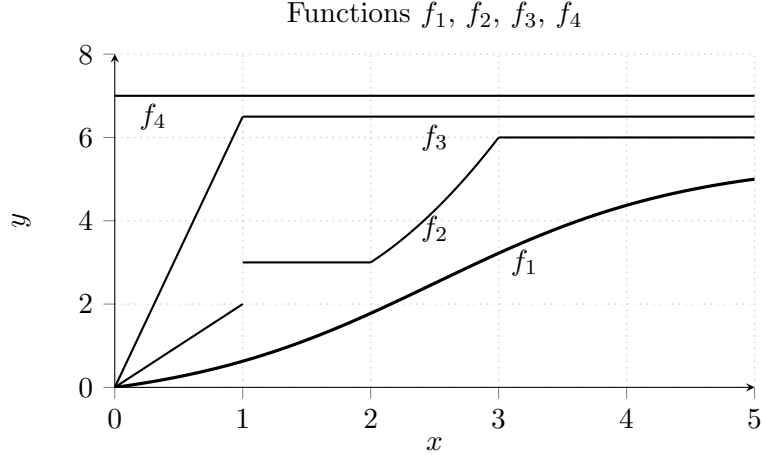
\begin{figure}[htbp]
\centering
\begin{tikzpicture}
\begin{axis}[
    width=10cm,
    height=6cm,
    xmin=0, xmax=5,
    ymin=0, ymax=8,
    axis lines=left,
    xlabel={$x$},
    ylabel={$y$},
    grid=both,
    grid style={dotted,gray!60},
    tick align=outside,
    title={Functions $f_1$, $f_2$, $f_3$, $f_4$}
]

\addplot[
    domain=0:5,
    samples=200,
    very thick,
    color=black
]
{ 5.8931*( 1/(1 + exp(-(x-2.5))) - 0.07586 ) };

\node at (axis cs:3.2,3) {$f_1$};

\addplot[domain=0:1, samples=50, thick, color=black] {2*x};
\addplot[domain=1:2, samples=2, thick, color=black] {3};
\addplot[domain=2:3, samples=50, thick, color=black] {3 + (x-2)*x};
\addplot[domain=3:5, samples=2, thick, color=black] {6};

\node at (axis cs:2.5,3.8) {$f_2$};

\addplot[domain=0:1, samples=50, thick, color=black] {6.5*x};
\addplot[domain=1:5, samples=2, thick, color=black] {6.5};

\node at (axis cs:2.5,6) {$f_3$};

\addplot[domain=0:5, samples=2, thick, color=black] {7};

\node at (axis cs:0.3,6.5) {$f_4$};

\end{axis}
\end{tikzpicture}
\caption{Schematic illustration of four possible coordination-strength functions.
  Functions $f_1$, $f_2$ and $f_3$ satisfy Assumption~\ref{ass:C1}, whereas $f_4$ has a
  singular slope at the origin and is not admissible.}
  \label{fig:f1234}

\end{figure}

The four functions in Figure~\ref{fig:f1234} all satisfy condition \eqref{CS2} (that is,
$y = 0$ when $x = 0$ and $y > 0$ when $x > 0$). Among them, $f_1$ is a $C^1$ function, and
$f_2$ and $f_3$ are piecewise $C^1$ functions. By contrast, $f_4$ is not of class $C^1$, because
its derivative at zero would be infinite.

We can now interpret the control-theoretic meaning of the $C^1$ requirement. By \eqref{CS2},
\[
  g_{ij}(\|v_i - v_j\|) > 0
  \quad \text{whenever} \quad \|v_i - v_j\| > 0.
\]
This means that $g_{ij}$ is fully sensitive at $x = 0$: even an arbitrarily small difference
between $v_i$ and $v_j$ immediately elicits a nonzero coordinating influence. Such immediate
sensitivity is necessary for state coordination.

If, instead, the coordination rule were
\[
  g_{ij}(\|v_i - v_j\|) > 0
  \quad \text{only when} \quad \|v_i - v_j\| \ge m \quad (m \ne 0),
\]
then coordination between individuals $i$ and $j$ would start only after their difference reaches
$\|v_i - v_j\| = m$ and would also end at that threshold. In that case the coordination processor would stop while $v_i \ne v_j$, and consensus would be impossible.

Requiring $g_{ij}$ to be $C^1$ (or piecewise $C^1$) ensures both sufficient sensitivity at
$x = 0$ and a moderate, graded response rather than an abrupt jump. It rules out singular
behaviours and keeps the coupling law physically and biologically reasonable.

From a biological standpoint, an ``infinite-slope'' response at zero distance, as in $f_4$, is
unrealistic. Real animals do not instantaneously switch from no response to a fixed, full-strength
reaction when the difference between two states becomes merely nonzero. Sensing and motor
systems operate in a continuous, graded manner.

For these reasons, the functions $f_1$, $f_2$, and $f_3$ are considered admissible forms of
coordination-strength functions in this paper: they remain smooth enough to be biologically
plausible while sufficiently responsive to support the coordination dynamics analysed below.

\subsection{Directed Graphs and Their Basic Properties}

Animal behaviour is constrained by perception, and animal sensory modalities---such as vision,
hearing, and olfaction---are inherently directional. An individual typically perceives others more
accurately in some directions than in others, and information propagates from the sender to the
receiver, not necessarily in a symmetric fashion. As a result, communication and coordination
within an animal group naturally form directed interaction patterns, which are conveniently
represented by directed graphs.

We briefly recall the basic structural properties of directed graphs that will be essential for the
subsequent analysis.

Let $\mathbb{G} = \langle \mathbb{A}, \mathcal{E}\rangle$ be a directed graph with vertex set
$\mathbb{A}$ and edge set $\mathcal{E}$.

\begin{define}[Strongly connected components and basic vertex sets]\label{def:SCC_basic}
A subgraph of $\mathbb{G}$ is called \emph{strongly connected} if every pair of nodes in this
subgraph is connected by a (possibly indirect) directed path in both directions. A maximal
strongly connected subgraph is called a \emph{strongly connected component} (SCC). Every
vertex belongs to exactly one SCC. The vertex set of an SCC is called a \emph{basic vertex set}.
\end{define}

Biologically, an SCC corresponds to a subgroup of individuals that can influence one another
through chains of directed interactions.

\begin{define}[Condensation graph and independent SCC]\label{def:condensation}
The \emph{condensation graph} of $\mathbb{G}$ is obtained by contracting each SCC into a single
vertex and retaining the induced directed edges between SCCs. The condensation graph is always
a directed acyclic graph (DAG), and thus admits a natural hierarchical structure.

An SCC is called an \emph{independent strongly connected component} (independent SCC)
if it does not receive any directed path from vertices outside itself. All other SCCs are called
\emph{non-independent SCCs}. The vertex sets of independent (resp.\ non-independent) SCCs
are called \emph{independent} (resp.\ \emph{non-independent}) basic vertex sets.
\end{define}

From a biological perspective, an independent SCC is a subgroup whose members may influence
others but are not, in the long run, influenced by any other subgroup.


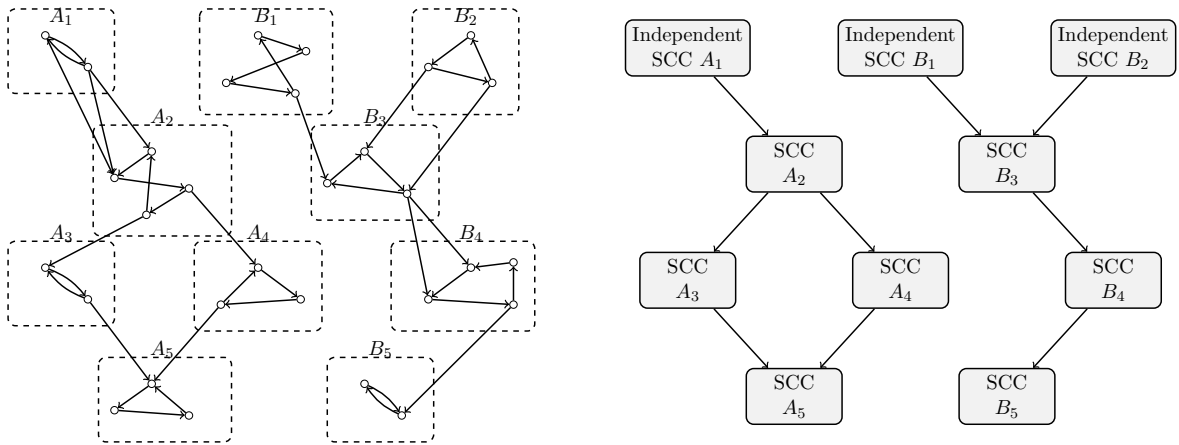
\begin{figure}[htbp]
\centering

\begin{minipage}{0.48\textwidth}
\centering
\scalebox{0.7}{
\begin{tikzpicture}[
    n/.style={circle, draw, inner sep=1pt, minimum size=4pt},
    SCCbox/.style={rounded corners, draw, thick, dashed},
    arr/.style={->, thick}
]


\node[n] (a1_1) at (-4,0) {};
\node[n] (a1_2) at (-3.2,-0.6) {};
\draw[arr] (a1_1) edge[bend left=15] (a1_2);
\draw[arr] (a1_2) edge[bend left=15] (a1_1);
\draw[SCCbox] (-4.7,0.5) rectangle (-2.7,-1.1);
\node at (-3.7,0.35) {$A_1$};

\node[n] (a2_1) at (0,0) {};
\node[n] (a2_2) at (0.9,-0.3) {};
\node[n] (a2_3) at (-0.6,-0.9) {};
\node[n] (a2_4) at (0.7,-1.1) {};
\foreach \i/\j in {1/2,2/3,3/4,4/1}{\draw[arr] (a2_\i) -- (a2_\j);}
\draw[SCCbox] (-1.1,0.5) rectangle (1.4,-1.5);
\node at (0.15,0.35) {$B_1$};

\node[n] (a3_1) at (4,0) {};
\node[n] (a3_2) at (3.2,-0.6) {};
\node[n] (a3_3) at (4.4,-0.9) {};
\draw[arr] (a3_1) -- (a3_2);
\draw[arr] (a3_2) -- (a3_3);
\draw[arr] (a3_3) -- (a3_1);
\draw[SCCbox] (2.9,0.5) rectangle (4.9,-1.5);
\node at (3.9,0.35) {$B_2$};


\node[n] (a4_1) at (-2,-2.2) {};
\node[n] (a4_2) at (-2.7,-2.7) {};
\node[n] (a4_3) at (-1.3,-2.9) {};
\node[n] (a4_4) at (-2.1,-3.4) {};
\foreach \i/\j in {1/2,2/3,3/4,4/1}{\draw[arr] (a4_\i) -- (a4_\j);}
\draw[SCCbox] (-3.1,-1.7) rectangle (-0.5,-3.8);
\node at (-1.8,-1.55) {$A_2$};

\node[n] (a5_1) at (2,-2.2) {};
\node[n] (a5_2) at (2.8,-3.0) {};
\node[n] (a5_3) at (1.3,-2.8) {};
\draw[arr] (a5_1) -- (a5_2);
\draw[arr] (a5_2) -- (a5_3);
\draw[arr] (a5_3) -- (a5_1);
\draw[SCCbox] (1.0,-1.7) rectangle (3.4,-3.5);
\node at (2.2,-1.55) {$B_3$};


\node[n] (a6_1) at (-4,-4.4) {};
\node[n] (a6_2) at (-3.2,-5.0) {};
\draw[arr] (a6_1) edge[bend left=15] (a6_2);
\draw[arr] (a6_2) edge[bend left=15] (a6_1);
\draw[SCCbox] (-4.7,-3.9) rectangle (-2.7,-5.5);
\node at (-3.7,-3.75) {$A_3$};

\node[n] (a7_1) at (0,-4.4) {};
\node[n] (a7_2) at (0.8,-5.0) {};
\node[n] (a7_3) at (-0.7,-5.1) {};
\draw[arr] (a7_1) -- (a7_2);
\draw[arr] (a7_2) -- (a7_3);
\draw[arr] (a7_3) -- (a7_1);
\draw[SCCbox] (-1.2,-3.9) rectangle (1.2,-5.6);
\node at (0,-3.75) {$A_4$};

\node[n] (a8_1) at (4,-4.4) {};
\node[n] (a8_2) at (3.2,-5.0) {};
\node[n] (a8_3) at (4.8,-5.1) {};
\node[n] (a8_4) at (4.8,-4.3) {};
\foreach \i/\j in {1/2,2/3,3/4,4/1}{\draw[arr] (a8_\i) -- (a8_\j);}
\draw[SCCbox] (2.5,-3.9) rectangle (5.2,-5.6);
\node at (4,-3.75) {$B_4$};


\node[n] (a9_1) at (-2,-6.6) {};
\node[n] (a9_2) at (-2.7,-7.1) {};
\node[n] (a9_3) at (-1.3,-7.2) {};
\draw[arr] (a9_1) -- (a9_2);
\draw[arr] (a9_2) -- (a9_3);
\draw[arr] (a9_3) -- (a9_1);
\draw[SCCbox] (-3.0,-6.1) rectangle (-0.5,-7.7);
\node at (-1.8,-5.95) {$A_5$};

\node[n] (a10_1) at (2,-6.6) {};
\node[n] (a10_2) at (2.7,-7.2) {};
\draw[arr] (a10_1) edge[bend left=15] (a10_2);
\draw[arr] (a10_2) edge[bend left=15] (a10_1);
\draw[SCCbox] (1.3,-6.1) rectangle (3.3,-7.7);
\node at (2.3,-5.95) {$B_5$};


\draw[arr] (a1_2) -- (a4_1);
\draw[arr] (a1_2) -- (a4_2);
\draw[arr] (a1_1) -- (a4_2);
\draw[arr] (a2_4) -- (a5_3);
\draw[arr] (a3_2) -- (a5_1);
\draw[arr] (a3_3) -- (a5_2);

\draw[arr] (a4_4) -- (a6_1);
\draw[arr] (a4_3) -- (a7_1);
\draw[arr] (a5_2) -- (a8_1);
\draw[arr] (a5_2) -- (a8_2);

\draw[arr] (a6_2) -- (a9_1);
\draw[arr] (a7_3) -- (a9_1);
\draw[arr] (a8_3) -- (a10_2);

\end{tikzpicture}
}

  \label{fig:condensation}

\end{minipage}
\hfill
\begin{minipage}{0.48\textwidth}
\centering
\scalebox{0.7}{
\begin{tikzpicture}[
    SCC/.style={rectangle, rounded corners, draw, thick,
                minimum width=1.8cm, minimum height=0.9cm,
                fill=gray!10, align=center},
    arr/.style={->, thick}
]

\node[SCC] (A1) at (-4,0)  {Independent\\SCC $A_1$};
\node[SCC] (A2) at (0,0)   {Independent\\SCC $B_1$};
\node[SCC] (A3) at (4,0)   {Independent\\SCC $B_2$};

\node[SCC] (A4) at (-2,-2.2) {SCC\\$A_2$};
\node[SCC] (A5) at (2,-2.2)  {SCC\\$B_3$};

\node[SCC] (A6) at (-4,-4.4) {SCC\\$A_3$};
\node[SCC] (A7) at (0,-4.4)  {SCC\\$A_4$};
\node[SCC] (A8) at (4,-4.4)  {SCC\\$B_4$};

\node[SCC] (A9)  at (-2,-6.6) {SCC\\$A_5$};
\node[SCC] (A10) at (2,-6.6)  {SCC\\$B_5$};

\draw[arr] (A1) -- (A4);
\draw[arr] (A2) -- (A5);
\draw[arr] (A3) -- (A5);

\draw[arr] (A4) -- (A6);
\draw[arr] (A4) -- (A7);
\draw[arr] (A5) -- (A8);

\draw[arr] (A6) -- (A9);
\draw[arr] (A7) -- (A9);
\draw[arr] (A8) -- (A10);

\end{tikzpicture}
}
\end{minipage}
  \caption{Original directed graph (left) and its condensation graph (right). Dashed boxes in
  the left panel indicate strongly connected components (SCCs). In the right panel, each SCC
  is contracted to a single node, yielding a directed acyclic condensation graph with a clear
  hierarchical structure.}
  \label{fig:condensation}

\end{figure}

If a directed graph has only one SCC, then the graph is said to be \emph{strongly connected}.
In this case, the entire graph forms a single independent SCC: all individuals are mutually
reachable (possibly through chains of intermediaries), and there is no strictly upstream or
downstream subgroup.

Figure~\ref{fig:condensation} illustrates two directed graphs together with their condensation
graphs. Graph~A contains a single independent strongly connected component, whereas Graph~B
contains two independent components. Biologically, this difference represents the contrast between
a group with one unique upstream ``source'' of influence and a group in which two such independent
``source'' subgroups coexist.

\subsection{Time-Varying Network}

In a time-varying interaction network, different pairs of individuals may spend very different total
amounts of time interacting. Such differences in accumulated interaction time inevitably affect
the long-term behaviour of the group. A convenient way to quantify this accumulated interaction
is to integrate the adjacency matrix of the time-varying network. Cao, Zheng and Zhou \cite{Cao2008,Cao2011}
introduced the following method for measuring edge lengths in time-varying networks and, in a
slightly different way, defined the notion of an $\infty$-adjoint graph.

\begin{define}[Edge-length measure and $\infty$-adjoint graph]\label{def:edge_length}
Let $\mathbb{G}(t) = \langle \mathbb{A}, \mathcal{E}(t), \mathbb{L}(t)\rangle$ be a time-varying
directed network, where $\mathbb{L}(t) = [\ell_{ij}(t)]$ is the adjacency matrix of $\mathbb{G}(t)$:
$\ell_{ij}(t) = 1$ if and only if $(a_i, a_j)_t \in \mathcal{E}(t)$, and $\ell_{ij}(t) = 0$ otherwise.

For each pair $(i,j)$, define the accumulated interaction time up to $T > 0$ as
\[
  L_{ij}(T) = \int_0^T \ell_{ij}(t) \, dt,
\]
and define the edge-length measure
\[
  \ell_{ij} = \int_0^\infty \ell_{ij}(t)\, dt
  = \lim_{T \to \infty} L_{ij}(T).
\]
The matrix
\[
  \mathbb{L}
  = \int_0^\infty \mathbb{L}(t)\, dt
  = \int_0^\infty
  \begin{pmatrix}
    \ell_{11}(t) & \cdots & \ell_{1n}(t) \\
    \vdots       & \ddots & \vdots       \\
    \ell_{n1}(t) & \cdots & \ell_{nn}(t)
  \end{pmatrix} dt
\]
is called the \emph{edge-length measure matrix} of $\mathbb{G}(t)$. The entry $\ell_{ij}$ is the
edge-length measure of the directed edge $(a_i, a_j)$.

The graph formed by all edges of $\mathbb{G}(t)$ whose edge-length measure is $\infty$ is
called the \emph{$\infty$-adjoint graph} of $\mathbb{G}(t)$ and is denoted by
\[
  \mathbb{G}_\infty = \langle \mathbb{A}, \mathcal{E}_\infty\rangle.
\]
\end{define}

The graph $\mathbb{G}_\infty$ extracts from the time-varying network $\mathbb{G}(t)$ a fixed
connectivity backbone by removing all edges $(a_i, a_j)$ whose edge-length measures are finite.
Edges with finite length represent interactions that are either genuinely temporary or, even if
recurrent, too weak in total duration to exert a significant long-term influence.

We now clarify the meaning of edges whose length is $\infty$ and explain their physical inter-
pretation. Assume that the time-varying network $\mathbb{G}(t)$ starts at $t=0$. For any $T>0$
and any edge $(a_i, a_j)$, the accumulated time $L_{ij}(T) = \int_0^T \ell_{ij}(t)\, dt$ satisfies
\[
  0 \le L_{ij}(T) \le T,
\]
since an edge cannot be active for longer than the length of the time interval itself. Suppose that,
for large $T$, we have
\[
  L_{ij}(T) \approx \frac{T}{2},
\]
meaning that the edge $(a_i, a_j)$ is active for roughly half of the time on $[0,T]$ and inactive
for the other half. Then, as $T \to \infty$, we obtain $L_{ij}(T) \to \infty$ and therefore
$\ell_{ij} = \int_0^\infty \ell_{ij}(t)\, dt = \infty$.

In terms of limits, the condition $\ell_{ij} = \infty$ simply states that, no matter how far we
go into the future, the edge $(a_i, a_j)$ continues to reappear with non-negligible total duration.
Equivalently, the interaction from $a_j$ to $a_i$ never fades away in the long run: its effect
remains integrable and has strictly positive total measure. The $\infty$-adjoint graph
$\mathbb{G}_\infty$ therefore consists exactly of those edges that keep reappearing indefinitely
and whose cumulative influence does not vanish.

For any $(a_i, a_j) \in \mathcal{E}_\infty$, consider the collection of time intervals
\begin{equation}\label{eq:I_active}
  \mathcal{I}(a_i, a_j)
  = \bigl\{
    (t_1,t_2), (t_3,t_4), (t_5,t_6), \dots
  \bigr\},
\end{equation}
which denote periods during which the directed interaction from $a_j$ to $a_i$ is active, and let
\begin{equation}\label{eq:I_inactive}
  \mathcal{I}'(a_i, a_j)
  = \bigl\{
    (t_2,t_3), (t_4,t_5), (t_6,t_7), \dots
  \bigr\},
\end{equation}
denote the complementary inactive intervals.


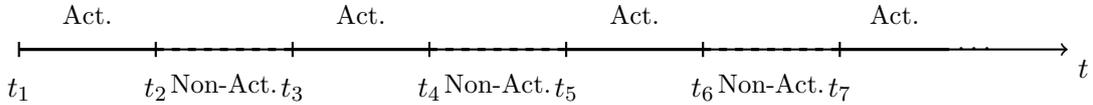
\begin{figure}[htbp]
\centering
\begin{tikzpicture}[x=1.2cm,y=1cm]

\draw[thick,->] (0,0) -- (11.5,0);
\node[below right] at (11.5,0) {$t$};

\foreach \x/\label in {0/t_1,1.5/t_2,3/t_3,4.5/t_4,6/t_5,7.5/t_6,9/t_7}
{
    \draw[thick] (\x,0.1) -- (\x,-0.1)
        node[below=4pt] {\(\label\)};
}

\node at (10.5,0) {$\cdots$};

\draw[very thick] (0,0) -- (1.5,0);
\draw[very thick] (3,0) -- (4.5,0);
\draw[very thick] (6,0) -- (7.5,0);
\draw[very thick] (9,0) -- (10.2,0);

\draw[very thick,dashed] (1.5,0) -- (3,0);
\draw[very thick,dashed] (4.5,0) -- (6,0);
\draw[very thick,dashed] (7.5,0) -- (9,0);

\node[above] at (0.75,0.2) {\small Act.};
\node[above] at (3.75,0.2) {\small Act.};
\node[above] at (6.75,0.2) {\small Act.};
\node[above] at (9.6,0.2) {\small Act.};

\node[below] at (2.25,-0.2) {\small Non-Act.};
\node[below] at (5.25,-0.2) {\small Non-Act.};
\node[below] at (8.25,-0.2) {\small Non-Act.};

\end{tikzpicture}

  \caption{Alternating intervals during which $a_i$ is affected by $a_j$ (Act.) or not affected
  (Non-Act.); the pattern extends indefinitely in time.}
  \label{fig:active_inactive}

\end{figure}

\begin{assump}\label{ass:stationary_intervals}
For each $(a_i, a_j) \in \mathcal{E}_\infty$ of the graph $\mathbb{G}_\infty$, the active and inactive
intervals satisfy:
\begin{align}
  \lim_{k\to\infty} |(t_k, t_{k+1})| &\ne 0,
  && (t_k, t_{k+1}) \in \mathcal{I}(a_i,a_j), \label{eq:active_nonzero}\\
  \lim_{k\to\infty} |(t_{k+1}, t_{k+2})| &\ne \infty,
  && (t_{k+1}, t_{k+2}) \in \mathcal{I}'(a_i,a_j). \label{eq:inactive_finite}
\end{align}
\end{assump}

If $\mathbb{G}_\infty$ satisfies Assumption~\ref{ass:stationary_intervals}, then there exist constants
$\tau_{\min} > 0$ and $\tau_B > 0$ such that for any $(a_i,a_j) \in \mathcal{E}_\infty$:
\begin{align}
  |(t_k, t_{k+1})| &\ge \tau_{\min},
  && (t_k, t_{k+1}) \in \mathcal{I}(a_i,a_j), \label{eq:tau_min}\\
  |(t_{k+1}, t_{k+2})| &\le \tau_B,
  && (t_{k+1}, t_{k+2}) \in \mathcal{I}'(a_i,a_j). \label{eq:tau_B}
\end{align}

Assumption~\ref{ass:stationary_intervals} ensures that the long-term influence encoded by the
$\infty$-adjoint graph $\mathbb{G}_\infty$ is not produced by pathological interaction patterns.
In particular, it prevents the active intervals from shrinking to zero length and excludes infinitely
long inactive gaps. Each edge in $\mathcal{E}_\infty$ therefore represents a genuinely persistent,
non-negligible interaction: it recurs throughout time, with a uniformly bounded minimal active
duration and a uniformly bounded maximal inactive interval. In this sense, the connectivity
structure captured by $\mathbb{G}_\infty$ reflects an enduring interaction relationship, rather
than one generated by rare or vanishingly weak events.

\begin{define}[Stationary time-varying network]\label{def:stationary_network}
A time-varying directed network $\mathbb{G}(t)$ is called a \emph{stationary time-varying
network} if its $\infty$-adjoint graph $\mathbb{G}_\infty$ satisfies Assumption~\ref{ass:stationary_intervals}.
\end{define}

In such cases, $\mathbb{G}_\infty$ encodes a fixed and persistent connectivity backbone that
remains invariant over time, even though the instantaneous interaction network $\mathbb{G}(t)$
may change continuously or discontinuously. This setting naturally describes real animal groups,
in which individuals continuously change their momentary neighbours but maintain stable topo-
logical neighbourhoods over longer time scales, a phenomenon documented by Ballerini
et al.\ \cite{Ballerini2008}.

\section{Core Theoretical Result}

The full mathematical analysis involves a substantial amount of notation, derivations and proofs.
These will be presented in Part~\ref{part:math}. In the present section we only state the core
theoretical result and explain its biological meaning.

\subsection{Main Theorem}

Under the framework of stationary time-varying networks introduced above, we obtain the
following necessary and sufficient condition for the velocity coordination system~\eqref{CS1} to
reach state consensus (Theorem~\ref{thm:nec_suff} below).

\paragraph{Main theorem (informal).}
\emph{Assume that the time-varying communication network $\mathbb{G}(t)$ is a stationary
time-varying network. Then all individuals in the velocity coordination system~\eqref{CS1}
reach state consensus if and only if the $\infty$-adjoint graph $\mathbb{G}_\infty$ of the
communication network contains a unique independent strongly connected component (SCC).}


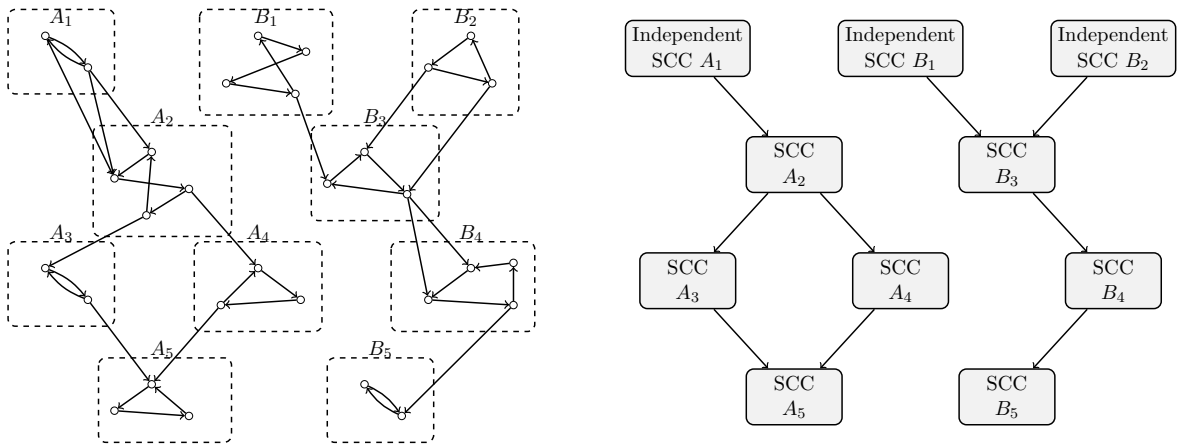
\begin{figure}[htbp]
\centering

\begin{minipage}{0.48\textwidth}
\centering
\scalebox{0.7}{
\begin{tikzpicture}[
    n/.style={circle, draw, inner sep=1pt, minimum size=4pt},
    SCCbox/.style={rounded corners, draw, thick, dashed},
    arr/.style={->, thick}
]


\node[n] (a1_1) at (-4,0) {};
\node[n] (a1_2) at (-3.2,-0.6) {};
\draw[arr] (a1_1) edge[bend left=15] (a1_2);
\draw[arr] (a1_2) edge[bend left=15] (a1_1);
\draw[SCCbox] (-4.7,0.5) rectangle (-2.7,-1.1);
\node at (-3.7,0.35) {$A_1$};

\node[n] (a2_1) at (0,0) {};
\node[n] (a2_2) at (0.9,-0.3) {};
\node[n] (a2_3) at (-0.6,-0.9) {};
\node[n] (a2_4) at (0.7,-1.1) {};
\foreach \i/\j in {1/2,2/3,3/4,4/1}{\draw[arr] (a2_\i) -- (a2_\j);}
\draw[SCCbox] (-1.1,0.5) rectangle (1.4,-1.5);
\node at (0.15,0.35) {$B_1$};

\node[n] (a3_1) at (4,0) {};
\node[n] (a3_2) at (3.2,-0.6) {};
\node[n] (a3_3) at (4.4,-0.9) {};
\draw[arr] (a3_1) -- (a3_2);
\draw[arr] (a3_2) -- (a3_3);
\draw[arr] (a3_3) -- (a3_1);
\draw[SCCbox] (2.9,0.5) rectangle (4.9,-1.5);
\node at (3.9,0.35) {$B_2$};


\node[n] (a4_1) at (-2,-2.2) {};
\node[n] (a4_2) at (-2.7,-2.7) {};
\node[n] (a4_3) at (-1.3,-2.9) {};
\node[n] (a4_4) at (-2.1,-3.4) {};
\foreach \i/\j in {1/2,2/3,3/4,4/1}{\draw[arr] (a4_\i) -- (a4_\j);}
\draw[SCCbox] (-3.1,-1.7) rectangle (-0.5,-3.8);
\node at (-1.8,-1.55) {$A_2$};

\node[n] (a5_1) at (2,-2.2) {};
\node[n] (a5_2) at (2.8,-3.0) {};
\node[n] (a5_3) at (1.3,-2.8) {};
\draw[arr] (a5_1) -- (a5_2);
\draw[arr] (a5_2) -- (a5_3);
\draw[arr] (a5_3) -- (a5_1);
\draw[SCCbox] (1.0,-1.7) rectangle (3.4,-3.5);
\node at (2.2,-1.55) {$B_3$};


\node[n] (a6_1) at (-4,-4.4) {};
\node[n] (a6_2) at (-3.2,-5.0) {};
\draw[arr] (a6_1) edge[bend left=15] (a6_2);
\draw[arr] (a6_2) edge[bend left=15] (a6_1);
\draw[SCCbox] (-4.7,-3.9) rectangle (-2.7,-5.5);
\node at (-3.7,-3.75) {$A_3$};

\node[n] (a7_1) at (0,-4.4) {};
\node[n] (a7_2) at (0.8,-5.0) {};
\node[n] (a7_3) at (-0.7,-5.1) {};
\draw[arr] (a7_1) -- (a7_2);
\draw[arr] (a7_2) -- (a7_3);
\draw[arr] (a7_3) -- (a7_1);
\draw[SCCbox] (-1.2,-3.9) rectangle (1.2,-5.6);
\node at (0,-3.75) {$A_4$};

\node[n] (a8_1) at (4,-4.4) {};
\node[n] (a8_2) at (3.2,-5.0) {};
\node[n] (a8_3) at (4.8,-5.1) {};
\node[n] (a8_4) at (4.8,-4.3) {};
\foreach \i/\j in {1/2,2/3,3/4,4/1}{\draw[arr] (a8_\i) -- (a8_\j);}
\draw[SCCbox] (2.5,-3.9) rectangle (5.2,-5.6);
\node at (4,-3.75) {$B_4$};


\node[n] (a9_1) at (-2,-6.6) {};
\node[n] (a9_2) at (-2.7,-7.1) {};
\node[n] (a9_3) at (-1.3,-7.2) {};
\draw[arr] (a9_1) -- (a9_2);
\draw[arr] (a9_2) -- (a9_3);
\draw[arr] (a9_3) -- (a9_1);
\draw[SCCbox] (-3.0,-6.1) rectangle (-0.5,-7.7);
\node at (-1.8,-5.95) {$A_5$};

\node[n] (a10_1) at (2,-6.6) {};
\node[n] (a10_2) at (2.7,-7.2) {};
\draw[arr] (a10_1) edge[bend left=15] (a10_2);
\draw[arr] (a10_2) edge[bend left=15] (a10_1);
\draw[SCCbox] (1.3,-6.1) rectangle (3.3,-7.7);
\node at (2.3,-5.95) {$B_5$};


\draw[arr] (a1_2) -- (a4_1);
\draw[arr] (a1_2) -- (a4_2);
\draw[arr] (a1_1) -- (a4_2);
\draw[arr] (a2_4) -- (a5_3);
\draw[arr] (a3_2) -- (a5_1);
\draw[arr] (a3_3) -- (a5_2);

\draw[arr] (a4_4) -- (a6_1);
\draw[arr] (a4_3) -- (a7_1);
\draw[arr] (a5_2) -- (a8_1);
\draw[arr] (a5_2) -- (a8_2);

\draw[arr] (a6_2) -- (a9_1);
\draw[arr] (a7_3) -- (a9_1);
\draw[arr] (a8_3) -- (a10_2);

\end{tikzpicture}
}
\end{minipage}
\hfill
\begin{minipage}{0.48\textwidth}
\centering
\scalebox{0.7}{
\begin{tikzpicture}[
    SCC/.style={rectangle, rounded corners, draw, thick,
                minimum width=1.8cm, minimum height=0.9cm,
                fill=gray!10, align=center},
    arr/.style={->, thick}
]

\node[SCC] (A1) at (-4,0)  {Independent\\SCC $A_1$};
\node[SCC] (A2) at (0,0)   {Independent\\SCC $B_1$};
\node[SCC] (A3) at (4,0)   {Independent\\SCC $B_2$};

\node[SCC] (A4) at (-2,-2.2) {SCC\\$A_2$};
\node[SCC] (A5) at (2,-2.2)  {SCC\\$B_3$};

\node[SCC] (A6) at (-4,-4.4) {SCC\\$A_3$};
\node[SCC] (A7) at (0,-4.4)  {SCC\\$A_4$};
\node[SCC] (A8) at (4,-4.4)  {SCC\\$B_4$};

\node[SCC] (A9)  at (-2,-6.6) {SCC\\$A_5$};
\node[SCC] (A10) at (2,-6.6)  {SCC\\$B_5$};

\draw[arr] (A1) -- (A4);
\draw[arr] (A2) -- (A5);
\draw[arr] (A3) -- (A5);

\draw[arr] (A4) -- (A6);
\draw[arr] (A4) -- (A7);
\draw[arr] (A5) -- (A8);

\draw[arr] (A6) -- (A9);
\draw[arr] (A7) -- (A9);
\draw[arr] (A8) -- (A10);

\end{tikzpicture}
}
\end{minipage}

  \caption{Original directed graphs (left) and their condensation graphs (right). In Panel A, the
  $\infty$-adjoint graph has a single independent SCC (A1). In Panel B, there are two such
  components (B1 and B2).}
  \label{fig:single_vs_multiple_SCC}

\end{figure}

In graph-theoretic terms, this condition says that the long-term interaction structure of the
group contains a single ``top-level'' subgroup that is not persistently influenced by any other
subgroup, whereas all other subgroups are, directly or indirectly, influenced by it. In biological
terms, this subgroup is the \emph{dominant subgroup}: its members mutually influence one another,
but are not persistently driven by anyone else in the long run. The state reached by this dominant
subgroup ultimately determines the common velocity of the entire group.

Figure~\ref{fig:single_vs_multiple_SCC} illustrates two typical directed-network structures.
In the first case (Panel~A), the $\infty$-adjoint graph $\mathbb{G}_\infty$ has a single independent
strongly connected component; in the second case (Panel~B), it has two. According to the main
theorem, the velocity coordination system reaches consensus only in the former case, where the
independent SCC is unique.

\subsection{Characteristics of the Consensus State}

We now consider the case in which $\mathbb{G}_\infty$ contains a unique independent strongly
connected component. Then, by Theorem~\ref{thm:nec_suff}, all individuals in the system
eventually reach consensus.

In Figure~\ref{fig:single_vs_multiple_SCC}A, this independent strongly connected component
is labelled A1. By definition, A1 does not receive persistent influence from any other component,
and its nodes influence one another through directed paths. Consequently, the individuals in A1
converge to a common state purely via their internal interactions. All other components (A2--A5)
are subordinate: they receive directed paths from A1, and therefore their states are gradually
driven towards the state reached by A1. Thus the final consensus of the entire system is determined
by the state to which the independent strongly connected component A1 converges.

Biologically, A1 is the dominant subgroup of the group: its members co-determine the final
velocity of the group, whereas all other individuals eventually follow.

\subsection{General Stability}

We next discuss the stability of a consensus state once it has formed.

The same structural condition applies not only to the formation of consensus, but also to its
maintenance. As long as the $\infty$-adjoint graph $\mathbb{G}_\infty$ remains unchanged, any
perturbation to the system has one of two outcomes:
\begin{itemize}
  \item If the dominant subgroup (the independent SCC) is not affected, then the system relaxes
  back to the original consensus state.
  \item If the dominant subgroup is affected, then the system converges to a new consensus state,
  again determined solely by the asymptotic state of this dominant subgroup.
  \end{itemize}

It is important to emphasize that the dominant subgroup is not defined by spatial location
(e.g., front-line or edge individuals), but by topology: it is the set of individuals that form the
unique independent strongly connected component in $\mathbb{G}_\infty$, i.e., the subgroup
whose long-term influence on the rest of the network is unidirectional.

\subsection{Stability Under Structural Changes}

A natural question is what happens if the structure of $\mathbb{G}_\infty$ itself changes over
time.

Theorem~\ref{thm:nec_suff} is a structural result: for any given stationary time-varying
network, it tells us under which connectivity pattern the system will exhibit ordered behaviour
(consensus). If the time-varying network evolves and later settles into a new stationary regime,
thereby defining a new $\infty$-adjoint graph $\mathbb{G}'_\infty$, then the same theorem applies
to this new structure.

In particular, if the new $\mathbb{G}'_\infty$ still contains a unique independent strongly
connected component, then the system will again reach consensus under the new structure. In
words:
\begin{quote}
  As long as the $\infty$-adjoint graph contains a unique independent strongly connected
  component, the final consensus state of the system is completely determined by this
  component. The detailed behaviour of subordinate non-independent components does not
  affect the final outcome.
\end{quote}

Under this paradigm, subordinate SCCs may undergo a wide range of internal changes---merg-
ing, splitting, reorganization, temporary disconnection, entry or exit of individuals---and may
even be drastically rearranged by environmental perturbations (predator attacks, local compres-
sion, sudden turns, etc.). However, as long as these changes do not create an additional independent
SCC, we have:
\begin{quote}
  No matter how non-core components change, they cannot alter the final consensus velocity of
  the group.
\end{quote}

The reason is straightforward: subordinate components do not exert a persistent return influence
on the independent SCC, whereas the latter continuously drives the former. Consequently, the full
system is always attracted to the consensus state set by the independent SCC.

\subsection{Correspondence with Empirical Data}

This theoretical picture aligns well with empirical observations of natural animal groups. The
STARFLAG project \cite{Ballerini2008} reconstructed three-dimensional trajectories of starling
flocks containing up to tens of thousands of individuals. Their analyses show that:
\begin{enumerate}
  \item Interaction rules are topological rather than metric: individuals align with a fixed number
  of nearest neighbours in the topological sense.
  \item Some neighbour relationships persist over long time intervals, whereas others occur only
  occasionally.
  \item These persistent relationships correspond to cumulative directional influences, and have
  decisive impact on group-level direction and velocity coordination.
\end{enumerate}

This empirical picture matches precisely the definition of the $\infty$-adjoint graph: persistent
interaction relationships correspond to edges in $\mathbb{G}_\infty$, and the subgroup formed by
such edges is the independent SCC (dominant subgroup) in our theory. Moreover, STARFLAG
data indicate that when a flock turns abruptly or is disturbed by predators, the instantaneous
topological structure may be rearranged; yet as long as there remains a single persistent topological
influence subgroup, the flock eventually recovers unified motion. Only when multiple persistent
topological influence subgroups emerge does the flock exhibit stable splitting or long-lasting
multi-subgroup structure.

\subsection{Formation of Large-Scale Moving Groups}

We now turn to the question of how large moving groups form in nature, and summarize the
associated behavioural mechanisms in a way that can be directly compared with empirical data.

A key point is that large aggregations are not formed by sparse individuals gradually cluster-
ing during motion. Instead, large groups emerge when a sufficiently dense population is almost
simultaneously triggered into coordinated dynamics. Multiple empirical studies support this view:
density-threshold experiments on locusts, sudden collective take-offs in roosting birds, synchronous
escape waves in fish schools, and the STARFLAG reconstructions of starling flocks. These studies
consistently indicate that the crucial precondition for coordination is the formation of a stable
topological neighbourhood structure under high-density conditions. Environmental or predatory
stimuli (light changes, acoustic shocks, approaching predators, etc.) merely activate this pre-
existing structure within a short time window.

Thus, large groups are not ``built'' during motion. Their internal organization is already present
during stationary or low-speed phases; when triggered, the group simply synchronizes rapidly into
a coordinated state. The coupling between high-density topological structure and fast activation
is the fundamental mechanism by which large animal groups form.

On this basis, we propose the following rules for group formation. They are partly supported
by existing behavioural observations, partly by reanalyses of available data, and partly by biolog-
ically reasonable inferences drawn from the assumptions and theoretical results of this paper.

\begin{enumerate}
  \item \textbf{Opportunistic formation.} Large groups appear in regions where individual density
  is already high, and multiple individuals (or subgroups) are activated simultaneously or nearly
  simultaneously. Formation thus occurs in a compact, high-density region within a short activation
  window.

  \item \textbf{Preparatory structure.} Before a large group forms, multiple moderate-sized sub-
  groups already exist in space-time. Because these subgroups are of moderate size and are
  ``primed'' for activation, they can rapidly achieve internal alignment once triggered.

  \item \textbf{Universality of the velocity-coordination rule.} When the trajectories of different
  subgroups intersect in space-time, subgroup merging occurs. The same velocity-coordination rule
  governs not only the formation and maintenance of groups of various sizes, but also the merging
  of subgroups. In this regime, effective interactions occur primarily between subgroups, and are
  transmitted through the internal coordination networks of each subgroup.

  \item \textbf{Directional-neighbour rule.} An individual's topological neighbours are distributed
  by direction. In particular, edge individuals must allocate attention both to other group members
  and to the external environment; their attentional field is approximately omnidirectional.

  \item \textbf{Inertial rule.} Once an individual has aligned its velocity with the group, it ceases
  to change its motion state unless perturbed. Equivalently, the group retains a ``trace'' of its most
  recent coordination event: after alignment, individuals move ballistically until a new coordination
  episode occurs.
\end{enumerate}

Rules (1) and (2) are strongly supported by observations and experiments, and represent the
opportunistic conditions under which large groups can form. We now explain their mechanistic
necessity.

When two individuals move with different directions, their mutual distance will grow over time
unless (a) coordination is distance-independent, or (b) the coordination is so strong that their
distance never exceeds the perceptual range (beyond which coordination is lost). Real animals
do not possess unlimited perceptual range or unlimited mechanical power. They therefore solve
this difficulty by pre-alignment at high density before activation, followed by nearly simultaneous
triggering. After activation, only relatively small further adjustments are needed for the group to
become aligned.

The motion groups formed under rules (1)--(2) are referred to as \emph{initial motion groups}.
There may be multiple such groups: they are initially separated in space-time, move with similar
but not identical velocities, and are mutually independent. After some time, two such groups may
enter a regime of spatiotemporal intersection, i.e., they are about to ``collide'' in space.

Our conclusion is that no additional behavioural rule is required beyond velocity coordination
itself. From basic relative-motion geometry, if two bodies have identical velocity vectors, their
relative position remains constant and they cannot collide; if their velocity vectors differ, their
relative position changes over time and may lead to intersection.


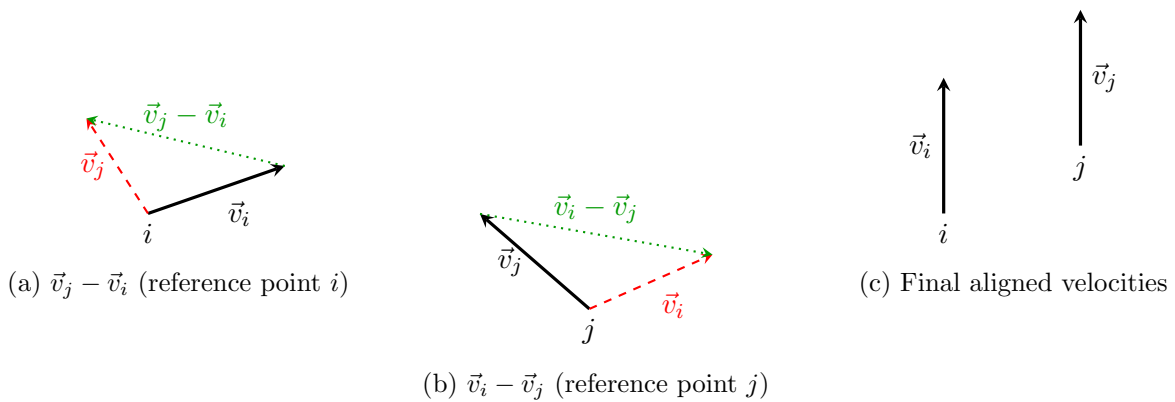
\begin{figure}[htbp]
\centering

\begin{subfigure}[b]{0.31\textwidth}
\centering
\begin{tikzpicture}[scale=0.9,>=stealth]

\tikzset{
  vmain/.style={very thick,->},
  vtrans/.style={thick,->,red,dashed},
  vdiff/.style={thick,->,green!60!black,dotted}
}

\node[below] at (0,0) {$i$};

\draw[vmain] (0,0) -- (2,0.7)
  node[midway,below right] {$\vec v_i$};

\draw[vtrans] (0,0) -- (-0.9,1.4)
  node[midway,left] {$\vec v_j$};

\draw[vdiff] (2,0.7) -- (-0.9,1.4)
  node[midway,above] {$\vec v_j - \vec v_i$};

\end{tikzpicture}
\caption{$\vec v_j - \vec v_i$ (reference point $i$)}
\end{subfigure}
\hfill
\begin{subfigure}[j]{0.31\textwidth}
\centering
\begin{tikzpicture}[scale=0.9,>=stealth]

\tikzset{
  vmain/.style={very thick,->},
  vtrans/.style={thick,->,red,dashed},
  vdiff/.style={thick,->,green!60!black,dotted}
}

\node[below] at (0,0) {$j$};

\draw[vmain] (0,0) -- (-1.6,1.4)
  node[midway,left] {$\vec v_j$};

\draw[vtrans] (0,0) -- (1.8,0.8)
  node[midway,below right] {$\vec v_i$};

\draw[vdiff] (-1.6,1.4) -- (1.8,0.8)
  node[midway,above] {$\vec v_i - \vec v_j$};

\end{tikzpicture}
\caption{$\vec v_i - \vec v_j$ (reference point $j$)}
\end{subfigure}
\hfill
\begin{subfigure}[b]{0.31\textwidth}
\centering
\begin{tikzpicture}[scale=0.9,>=stealth]

\tikzset{
  vmain/.style={very thick,->}
}

\node[below] at (0,0) {$i$};
\draw[vmain] (0,0) -- (0,2)
  node[midway,left] {$\vec v_i$};

\node[below] at (2,1) {$j$};
\draw[vmain] (2,1) -- (2,3)
  node[midway,right] {$\vec v_j$};

\end{tikzpicture}
\caption{Final aligned velocities}
\end{subfigure}

  \caption{Construction of velocity-difference vectors by translation of $\bm{v}_i$ and $\bm{v}_j$,
  and the resulting final alignment.}
  \label{fig:relative_vel}

\end{figure}

Figure~\ref{fig:relative_vel} illustrates how two moving individuals avoid an imminent collision
through velocity coordination and ultimately form a single moving group. According to the
velocity-coordination law~\eqref{CS1} and the condition on the coordination-strength function
\eqref{CS2}, individuals $i$ and $j$ generate accelerations along these difference vectors until
$\|v_i - v_j\| \to 0$. The final state is one in which their velocities are identical in both magnitude
and direction.

In most species, variation in speed (velocity magnitude) among individuals is modest; large
velocity differences are dominated by directional discrepancies. Even when speeds are nearly equal,
different directions can produce large vectors $\bm{v}_j - \bm{v}_i$, so collision-avoidance
manoeuvres driven by directional differences tend to be rapid and strongly directional. This
mechanism generalizes directly from pairs of individuals to interacting subgroups: when two sub-
groups enter each other's neighbourhood, some individuals form cross-links, and the resulting
velocity coordination propagates through each subgroup's internal interaction network.

In most cases, two initial groups that begin to interact under these rules will merge into a single
group, although fragile ``tearing'' events are also possible. Whether merging or splitting occurs
depends on factors such as:
\begin{itemize}
  \item the distance and velocity angle when interaction begins;
  \item the locations of the first cross-links (front, middle, or tail);
  \item whether cross-links are mutual or unidirectional;
  \item the relative sizes of the two groups (both small/medium, one large and one small, both large);
  \item the structural robustness and plasticity of the internal networks.
\end{itemize}

As a concrete example, consider a \emph{rear-end braking} mode of group merging. Multiple
initial motion groups appear after activation, moving in roughly the same direction with similar
speeds. Because migration directionality is strong, these groups do not diverge rapidly. Small
differences in velocity, perturbations and noise, however, guarantee that one group eventually
becomes slower than the others. Trailing groups then catch up with it, and to avoid collision they
begin velocity coordination. In a typical scenario:
\begin{enumerate}
  \item The leading group (A) and the trailing group (B) interact at a small velocity angle.
  \item Front individuals of group~B coordinate with tail individuals of group~A.
  \item The coordination is effectively unidirectional: there are persistent edges from A to B in
  $\mathbb{G}_\infty$, but no persistent edges from B to A.
\end{enumerate}

Under these conditions, the cross-links attach group~B to group~A in the $\infty$-adjoint graph,
forming a new group that contains both subgroups and has a single independent SCC, namely the
original front line of A. By the main theorem, the new group maintains consensus, and its consensus
velocity coincides with that of the original dominant subgroup of A. As a result, all individuals
in the trailing group~B slow down. This is the mechanism of \emph{velocity inheritance}.

Note that the trailing group need not be directly behind the leading group: a rear-lateral
approach can increase the width and thickness of the tail region, thereby enlarging the ``capture
cross-section'' for absorbing trailing groups. Observational and empirical studies suggest that this
mechanism contributes to the characteristic ``narrow-front, wide-back; sparse-front, dense-back''
geometry of many large animal groups.

This leads to several empirically testable predictions. For example:
\begin{itemize}
  \item Comparing the speeds of groups of different sizes---especially very large groups versus
  small or very small groups (likely corresponding to initial groups)---should reveal that ultra-large
  groups move more slowly than the mean speed of initial groups in the population.
  \item Ultra-large groups should appear only when initial groups have sufficiently high density
  at activation, reflecting the opportunistic nature of large-group formation.
\end{itemize}

Finally, we return to the inertial rule. In our model, condition~\eqref{CS2} implies that once
$v_i = v_j$ for all neighbouring individuals, there is no further coordinating acceleration. The
group then moves inertially until a new perturbation or coordination episode occurs. This has two
corollaries:
\begin{itemize}
  \item At the instant when subgroups merge into a single group, individuals within each subgroup
  stop adjusting their motion state (unless perturbed), so inter-individual distances at that instant
  need not display strong regularity.
  \item Observing such ``frozen'' subgroup structure in large moving groups would suggest a history
  of subgroup merging and would support the picture of pure velocity-based (first-order) coordi-
  nation dynamics developed in this paper.
\end{itemize}

\section{Testable Predictions}

In this section we summarize several empirical predictions that follow directly from the theoretical
framework developed above. Each prediction includes its theoretical basis, its empirical implication,
and specific indicators that can be tested using STARFLAG-style three-dimensional data.

\subsection{Prediction 1: Network-Structure Hierarchy}

\textbf{Theoretical basis.} The main consensus theorem, which states that the final consensus
state is determined by the unique independent strongly connected component (SCC).

\textbf{Prediction.} The topological interaction network of any real animal group should
decompose into multiple SCCs. These SCCs must form a directed hierarchical structure:
\begin{itemize}
  \item the first layer contains a unique independent SCC;
  \item the second layer consists of SCCs depending on the first layer;
  \item the third layer depends on the second layer;
  \item and so on.
\end{itemize}
Spatially, this hierarchy should project along the movement direction as a ``front-dominant,
rear-subordinate'' structure.

\textbf{Empirical indicators.}
\begin{itemize}
  \item Reconstruct topological neighbour networks from STARFLAG-style 3D data.
  \item Identify SCCs and construct the condensation graph; verify the uniqueness of the top-level
  SCC.
  \item Test whether hierarchical layers align with the spatial order along the movement direction.
\end{itemize}

\subsection{Prediction 2: Inertia Rule and Distance Distributions}

\textbf{Theoretical basis.} First-order velocity-coordination dynamics together with the struc-
tural properties of the coordination-strength function.

\textbf{Prediction.}
\begin{itemize}
  \item Distances between individuals and their topological neighbours should exhibit a broad,
  non-structured distribution (consistent with STARFLAG).
  \item Distances between ``linking individuals'' that connect different SCC layers should also
  be broadly distributed, without forming any fixed geometric pattern.
\end{itemize}

\textbf{Relation to empirical findings.} STARFLAG has demonstrated that topological inter-
actions produce a broad distribution of individual--neighbour distances \cite{Ballerini2008}.
However, STARFLAG did not analyze cross-SCC distances. Thus, the second point constitutes
a novel empirical prediction uniquely arising from our theory.

\textbf{Empirical indicators.}
\begin{itemize}
  \item Identify individuals forming cross-SCC connections and measure the distances between
  them.
  \item Test whether these distances follow a broad, non-structured distribution.
\end{itemize}

\subsection{Prediction 3: Directional-Neighbor Structure}

\textbf{Theoretical basis.} Directional sensitivity of individuals and geometric prerequisites for
subgroup merging.

\textbf{Prediction.}
\begin{itemize}
  \item The angular distribution of topological neighbours should show detectable statistical
  structure across different directions.
  \item Edge individuals, lacking neighbours on one side, should exhibit reduced neighbour counts
  and a compensatory bias toward the opposite direction.
\end{itemize}

\textbf{Empirical indicators.}
\begin{itemize}
  \item Compute neighbour counts across angular sectors (left/right, front/back).
  \item Analyze edge individuals specifically for unilateral deficits and directional bias.
\end{itemize}

\subsection{Prediction 4: Reduced Speed of Large Groups}

\textbf{Theoretical basis.} Rear-end braking dynamics and the inheritance of velocity from the
independent SCC.

\textbf{Prediction.} Ultra-large coordinated groups should exhibit speeds significantly below
the mean speed of the initial population of small and medium groups, and the distribution should
show a left shift.

\textbf{Empirical indicators.}
\begin{itemize}
  \item Measure speed distributions of small and intermediate-sized groups.
  \item Measure the speed of large groups independently.
  \item Test for statistically significant left-shifted deviation relative to the initial mean.
\end{itemize}

\subsection{Prediction 5: Asymmetric Group Geometry}

\textbf{Theoretical basis.} Rear-end braking and tail-absorption mechanisms during subgroup
merging.

\textbf{Prediction.} Large groups should display a characteristic asymmetric geometry:
\begin{itemize}
  \item a narrow, low-density front;
  \item a wide, high-density rear;
  \item progressively expanding tail geometry with repeated merging events.
\end{itemize}

\textbf{Empirical indicators.}
\begin{itemize}
  \item Compute density and geometric cross-sections along the movement direction.
  \item At the SCC level, compare spatial extent and density patterns across subgroups.
\end{itemize}

\part{Mathematical Appendix}\label{part:math}

\section{Theoretical Derivation}

\subsection{Projection Theorem}

Let $\bm{r}_k$ be the unit vector along the $k$-th coordinate axis in $\mathbb{R}^m$
($k = 1,\dots,m$). The projection system of system~\eqref{CS1} onto the $k$-th coordinate axis
is
\begin{equation}\label{eq:projection}
  \dot{v}_i^{(k)}
  = \sum_{a_j \in N(t,a_i)} -g_{ij}(\|v_i - v_j\|)\,
  \cos(\bm{r}_{ij}, \bm{r}_k)
  = \sum_{a_j \in N(t,a_i)} -\kappa_{ij}^k
  \frac{v_i^{(k)} - v_j^{(k)}}{|v_i^{(k)} - v_j^{(k)}|},
\end{equation}
where $v_i^{(k)}$ denotes the $k$-th component of $v_i \in \mathbb{R}^m$, $\bm{r}_{ij}$ is the
unit vector in the direction $v_i - v_j$, and
\[
  \kappa_{ij}^k
  = g_{ij}(\|v_i - v_j\|)\,
    \frac{|v_i^{(k)} - v_j^{(k)}|}{\|v_i - v_j\|}
  = g_{ij}(\|v_i - v_j\|)\,|\cos(\bm{r}_{ij}, \bm{r}_k)|.
\]

The following result is established in \cite{JinZheng2011_affine,JinZheng2011_nonlinear,JinZheng2011_unified}.

\begin{prop}[Cooperativity of the projection system]\label{prop:projection}
A necessary and sufficient condition for the affine multi-agent system~\eqref{CS1} to be a
velocity coordination system is that its projection system~\eqref{eq:projection} onto any
coordinate axis in $\mathbb{R}^m$ is also a velocity coordination system. In particular, when
$g_{ij}$ satisfies \eqref{CS2}, the quantity $\kappa_{ij}^k$ satisfies
\[
  \kappa_{ij}^k
  =
  \begin{cases}
    0, & |v_i^{(k)} - v_j^{(k)}| = 0,\\[0.2em]
    > 0, & |v_i^{(k)} - v_j^{(k)}| > 0.
  \end{cases}
\]
\end{prop}

Hence the consensus problem in $\mathbb{R}^m$ can be reduced to consensus problems on
one-dimensional projections along coordinate axes, without loss of generality.

\subsection{Lyapunov Stability Analysis}

This section examines the Lyapunov stability of the affine velocity coordination system
\eqref{CS1}. This represents the most fundamental property of velocity coordination systems
and is independent of the underlying interaction network.

Let
\[
  V(t) = \{ v_i(t) \mid a_i \in \mathbb{A},\, t \ge 0 \} \subset \mathbb{R}^m
\]
denote the collection of individual states of system~\eqref{CS1} at time $t$.

\begin{define}[Minimal convex hull]\label{def:convex_hull}
A convex set $\Xi(t) \subset \mathbb{R}^m$ is called the \emph{minimal convex hull} of $V(t)$
if for any convex set $\Theta(t)$ satisfying $V(t) \subset \Theta(t) \subset \mathbb{R}^m$, it holds
that
\[
  V(t) \subset \Xi(t) \subset \Theta(t).
\]
\end{define}

The following results is established in {\cite{JinZheng2011_affine,JinZheng2011_nonlinear,JinZheng2011_unified}}.

\begin{thm}\label{thm:convex_nonexpanding}
For any $t \ge 0$ and any $\delta t > 0$, the minimal convex hull $\Xi(t)$ of the state set
$V(t) \subset \mathbb{R}^m$ of the affine velocity coordination system~\eqref{CS1} satisfies
\[
  \Xi(t + \delta t) \subset \Xi(t).
\]
\end{thm}

This implies that the state of the system, after being perturbed, will never leave the convex
region determined at the end of the perturbation. Consequently, the system is Lyapunov stable.

\begin{cor}\label{cor:Lyapunov}
The affine velocity coordination system~\eqref{CS1} is Lyapunov stable.
\end{cor}

By the Weierstrass theorem, we obtain:

\begin{cor}\label{cor:limit_convex_hull}
The limit of the minimal convex hull $\Xi(t)$ of the state set $V(t)$ of system~\ref{CS1} exists:
\[
  \Xi(\infty) = \lim_{t \to \infty} \Xi(t).
\]
\end{cor}

\subsection{Necessary and Sufficient Condition for Achieving Consensus}

According to Proposition~\ref{prop:projection}, the projection systems~\eqref{eq:projection}
of the velocity coordination system~\eqref{CS1} onto the coordinate axes of $\mathbb{R}^m$
are themselves velocity coordination systems. Since $g_{ij}$ is of class $C^1$, the function
$\kappa_{ij}^k = g_{ij}(\|v_i - v_j\|)\,|\cos(\bm{r}_{ij},\bm{r}_k)|$ is also of class $C^1$. Clearly,
the velocity coordination system~\eqref{CS1} achieves state consensus if and only if all its
projection systems~\eqref{eq:projection} on the coordinate axes of $\mathbb{R}^m$ achieve state
consensus. Therefore, it suffices to consider the case $v_i \in \mathbb{R}$ for $i=1,\dots,n$.

Let
\[
  V(t) = \{ v_i(t) \mid a_i \in \mathbb{A}\}, \quad
  \Delta(t) = [\underline{\Delta}(t), \overline{\Delta}(t)]
  = [\min V(t), \max V(t)].
\]
By Corollary~\ref{cor:limit_convex_hull}, the limit of $\Delta(t)$ exists. Denote
\[
  [\Delta_1, \Delta_2] = \lim_{t\to\infty} \Delta(t).
\]
Then, by Theorem~\ref{thm:convex_nonexpanding}, for any $\varepsilon>0$, there exists $t_k$
such that for all $t>t_k$,
\begin{equation}\label{eq:Delta_eps}
  v_i(t) \in (\Delta_1 - \varepsilon, \Delta_2 + \varepsilon),
  \quad a_i \in \mathbb{A}.
\end{equation}

We now state two lemmas (bounds on the dynamics and propagation of near-maximum states)
and then the sufficient condition for consensus.

\begin{lemma}\label{prop:upper_bound}
Under \eqref{eq:Delta_eps}, there exists a constant $\gamma>0$ such that for all $t\in[t_k,\infty)$,
\[
  \dot{v}_i(t) \le \gamma(\Delta_2 + \varepsilon - v_i(t)),
  \quad \forall a_i \in \mathbb{A}.
\]
\end{lemma}
\textbf{Proof:}~

Consider the interaction strength functions $g_{ij}(y)$ on the closed interval
$y\in[0,\Delta_2-\Delta_1+2\varepsilon]$. Define
\[
\gamma_{ij}
=
\max_{y\in[0,\Delta_2-\Delta_1+2\varepsilon]}
\frac{g_{ij}(y)}{y}.
\]
By the definition of the interaction strength function (\ref{CS2}),
$g_{ij}(y)\ge 0$ for all
$y\in[0,\Delta_2-\Delta_1+2\varepsilon]$.
Moreover, by \textbf{Assumption}~\ref{ass:C1}, the limit
\[
\lim_{y\to 0}\frac{g_{ij}(y)}{y}
\]
exists. Hence $\gamma_{ij}$ exists and satisfies $\gamma_{ij}>0$.

\[
\dot v_i
= \sum_{a_j\in N(t,a_i)}
   -\frac{g_{ij}(\|v_i-v_j\|)}{\|v_i-v_j\|}
    (v_i-v_j)
= \sum_{a_j\in N(t,a_i)}
   \frac{g_{ij}(\|v_i-v_j\|)}{\|v_i-v_j\|}(v_j-v_i)
\]
\[
\le
\sum_{a_j\in N(t,a_i)}
   \gamma_{ij}\bigl(\Delta_2+\varepsilon - v_i\bigr).
\]

Let
\[
\gamma'=
\max\{\gamma_{ij}\mid i,j\in\underline{n}\}.
\]
Then
\[
\dot v_i
\le
\sum_{a_j\in N(t,a_i)}
  \gamma_{ij}\bigl(\Delta_2+\varepsilon - v_i\bigr)
\le
\sum_{a_j\in N(t,a_i)}
  \gamma'\bigl(\Delta_2+\varepsilon - v_i\bigr)
\]
\[
\le
\sum_{\substack{j=1\\ j\ne i}}^{\,n}
  \gamma'\bigl(\Delta_2+\varepsilon - v_i\bigr)
= (n-1)\gamma'\,\bigl(\Delta_2+\varepsilon - v_i\bigr).
\]

Setting $\gamma=(n-1)\gamma'$, we obtain
\[
\dot v_i \le \gamma(\Delta_2+\varepsilon - v_i),
\]
as desired.

\hfill$\Box$

\begin{lemma}\label{lem:backwards}
For the velocity coordination system~\eqref{CS1}, let $\varepsilon>0$ and $\varepsilon'>0$
be arbitrarily small. If
\[
  v_i(t_k) \in (\Delta_2 - \varepsilon', \Delta_2 + \varepsilon),
\]
then for any $\tau>0$ and all $t \in [t_k - \tau, t_k]$,
\[
  v_i(t) \in (\Delta_2 - \varepsilon'', \Delta_2 + \varepsilon),
\]
where $\varepsilon''>0$ is an infinitesimal equivalent to $\varepsilon$ and $\varepsilon'$.
\end{lemma}

\textbf{Proof:}~

Assume that $v_i(t_k-\tau)<\Delta_2-\varepsilon'$ and that $a_i$ approaches
$v_i(t_k)=\Delta_2-\varepsilon'$ at time $t_k$. By
\textbf{lemma}~\ref{prop:upper_bound}, there exists a constant $\gamma>0$ such that
\[
\dot v_i < \gamma\bigl((\Delta_2+\varepsilon)-v_i\bigr).
\]
Hence,
\[
v_i(t_k-\tau)
>
(\Delta_2+\varepsilon)
   - (\varepsilon+\varepsilon') e^{\gamma\tau},
\]
that is,
\[
v_i(t_k-\tau)
\in
\bigl(\Delta_2+\varepsilon - (\varepsilon+\varepsilon')e^{\gamma\tau},\;
      \Delta_2+\varepsilon\bigr).
\]
Because $\tau>0$ and $\gamma>0$, we have
$(\varepsilon+\varepsilon')e^{\gamma\tau}-\varepsilon>0$.
Let
\[
\varepsilon'' = (\varepsilon+\varepsilon')e^{\gamma\tau}-\varepsilon.
\]
Then
\[
v_i(t_k-\tau)\in (\Delta_2-\varepsilon'',\, \Delta_2+\varepsilon).
\]

\hfill$\Box$

\begin{lemma}\label{lem:edge_prop}
Suppose that $\mathbb{G}(t)$ is a stationary time-varying network and that
$(a_i,a_j)_\infty \in \mathcal{E}_\infty$. If
\[
  v_i(t) \in (\Delta_2 - \varepsilon'', \Delta_2 + \varepsilon)
  \quad \text{for all } t \in [t_k - (\tau_B + \tau_{\min}), t_k],
\]
then, for $\varepsilon$ and $\varepsilon''$ sufficiently small, there exists some
$t'\in[t_k - (\tau_B + \tau_{\min}), t_k]$ such that
\[
  v_j(t') \in (\Delta_2 - \varepsilon''', \Delta_2 + \varepsilon),
\]
where $\varepsilon'''>0$ is an infinitesimal equivalent to $\varepsilon$ and $\varepsilon''$.
\end{lemma}
\textbf{Proof.~}

Suppose $(a_i,a_j)_\infty\in{\cal E}_\infty$, and that
\[
v_i(t)\in(\Delta_2-\varepsilon'',\,\Delta_2+\varepsilon)
\qquad\text{for all }t\in[t_k-(\tau_B+\tau_{min}),\,t_k].
\]

By the stationarity of $\mathbb{G}(t)$, there exists an interval
\[
[t'_k-\tau_{min},\,t'_k)\subset [t_k-(\tau_B+\tau_{min}),\,t_k]
\]
such that
\begin{equation}\label{lem2-00}
(a_i,a_j)_t \in {\cal E}(t)
\qquad\text{for all } t\in[t'_k-\tau_{min},\,t'_k).
\end{equation}

If there exists some
\[
t' \in [t'_k-\tau_{min},\,t'_k)
\]
such that
\[
v_j(t') \in (\Delta_2-\varepsilon'',\,\Delta_2+\varepsilon),
\]
the lemma is proved.
Hence, without loss of generality, assume that for all
$t\in[t'_k-\tau_{min},\,t'_k)$,
\begin{equation}\label{lem2-01}
v_j(t)\;\le\;\Delta_2-\varepsilon''' \;<\; \Delta_2-\varepsilon''.
\end{equation}

From
\begin{equation}\label{lem2-02}
\dot v_i(t)
=\sum_{a_s\in N(t,a_i)} -g_{is}\!\left(\|v_i-v_s\|\right)\vec r_{is}
=\sum_{a_s\in N(t,a_i)}
-g_{is}\!\left(\|v_i-v_s\|\right)
\frac{v_i-v_s}{\|v_i-v_s\|},
\end{equation}
and by Proposition~\ref{prop:projection}, it suffices to consider the scalar case
$v_i,x_s\in\mathbb{R}^1$.
Then \eqref{lem2-02} becomes
\begin{equation}\label{lem2-03}
\dot v_i
=\sum_{a_s\in N(t,a_i)} -g_{is}(\|v_i-v_s\|)
=f_{iL}+f_{iR},
\end{equation}
where
\[
f_{iL}=\sum_{a_l\in N(t,a_i),\,v_l<v_i} -g_{il}(\|v_i-v_l\|)\le 0,
\qquad
f_{iR}=\sum_{a_r\in N(t,a_i),\,v_r\ge v_i} g_{ir}(\|v_i-v_r\|)\ge 0.
\]
By \eqref{lem2-00} and \eqref{lem2-01}, at least one left neighbor $a_j$
acts on $a_i$ for all $t\in[t'_k-\tau_{min},\,t'_k)$, so
\begin{equation}\label{lem2-06}
f_{iL}
\le -\,g_{ij}\!\left(\|v_i-v_j\|\right).
\end{equation}

Define
\begin{equation}\label{lem2-04}
V(\varepsilon,\varepsilon'')
=\max\Big\{ g_{is}(y)\,\big|\,
0\le y\le \varepsilon+\varepsilon'',\; i\ne s\Big\} > 0.
\end{equation}

Since
$v_i(t)\in(\Delta_2-\varepsilon'',\,\Delta_2+\varepsilon)$
over the interval, and $v_r\ge v_i$, we have
$\|v_i-v_r\|\le \varepsilon+\varepsilon''$ and thus
\[
g_{ir}(\|v_i-v_r\|)\le V(\varepsilon,\varepsilon'').
\]
Hence
\begin{equation}\label{lem2-05}
f_{iR}\le (n-2)V(\varepsilon,\varepsilon'')
=:V_R(\varepsilon,\varepsilon''),
\end{equation}
and $V_R(\varepsilon,\varepsilon'')$ is an infinitesimal equivalent to
$\varepsilon,\varepsilon''$, i.e.,
\begin{equation}\label{lem2-04-1}
V_R(\varepsilon,\varepsilon'')\to 0
\quad\text{as}\quad
\varepsilon,\varepsilon''\to 0.
\end{equation}

Substituting \eqref{lem2-06} and \eqref{lem2-05} into \eqref{lem2-03}
gives
\[
\dot v_i
\le -g_{ij}(\|v_i-v_j\|)+V_R(\varepsilon,\varepsilon'').
\]

Integrating over $[t'_k-\tau_{min},\,t'_k)$ yields
\[
\int_{t'_k-\tau_{min}}^{t'_k} g_{ij}(\|v_i(t)-v_j(t)\|)\,dt
\le \varepsilon+\varepsilon'' + V_R(\varepsilon,\varepsilon'')\tau_{min}.
\]

By the mean value theorem, there exists
$t'\in[t'_k-\tau_{min},\,t'_k)$ such that
\begin{equation}\label{lem2-5}
g_{ij}(\|v_i(t')-v_j(t')\|)
\le \frac{\varepsilon+\varepsilon''}{\tau_{min}}
+V_R(\varepsilon,\varepsilon'').
\end{equation}

If $g_{ij}$ is not strictly monotone, let $\eta$ be the minimal local
extremum value of $g_{ij}$.
Choose $\eta'<\eta$ and $\delta'>0$ such that $g_{ij}(\delta')=\eta'$;
on $[0,\delta')$ the function is strictly increasing.
Since $(\varepsilon+\varepsilon'')/\tau_{min}+V_R(\varepsilon,\varepsilon'')$
is an infinitesimal, we may assume it is $\le \eta'$, so
\begin{equation}\label{lem2-6}
\|v_i(t')-v_j(t')\|
\le g_{ij}^{-1}\!\left(
\frac{\varepsilon+\varepsilon''}{\tau_{min}}
+V_R(\varepsilon,\varepsilon'')
\right).
\end{equation}

The right-hand side tends to $0$ as
$\varepsilon,\varepsilon''\to 0$.
Since
$\|v_i(t')-\Delta_2\|<\varepsilon+\varepsilon''$, we conclude
\[
\|v_j(t')-\Delta_2\|
\le (\varepsilon+\varepsilon'')
+ g_{ij}^{-1}\!\left(
\frac{\varepsilon+\varepsilon''}{\tau_{min}}
+V_R(\varepsilon,\varepsilon'')\right)
=\varepsilon'''.
\]

\hfill$\Box$

Using these lemmas, one propagates ``almost-maximum'' states backward along directed paths
in $\mathbb{G}_\infty$ and obtains the following result.

\begin{thm}[Sufficient condition]\label{thm:sufficient}
The velocity coordination system~\eqref{CS1} achieves state consensus for all individuals if:
\begin{enumerate}[(1)]
  \item $\mathbb{G}(t)$ is a stationary time-varying network; and
  \item the $\infty$-adjoint graph $\mathbb{G}_\infty = \langle \mathbb{A}, \mathcal{E}_\infty\rangle$
        has a unique independent strongly connected component.
\end{enumerate}
\end{thm}
\textbf{Proof.}

We consider the case $v_i\in\mathbb{R}^1$.
By \eqref{cor:limit_convex_hull}, for any arbitrarily small $\varepsilon>0$ there exists
$t_k$ such that, for all $t>t_k$,
\[
\begin{array}{c}
   v_i(t)\in(\Delta_1-\varepsilon,\ \Delta_2+\varepsilon),
   \qquad a_i\in\mathbb{A},
\end{array}
\]
where $[\Delta_1,\Delta_2]$ is the limit, as $t\to\infty$, of the minimal
convex cover
\[
\Delta(t)=[\underline{\Delta}(t),\overline{\Delta}(t)]
\]
of the state set of system \eqref{CS1}.

\medskip
\noindent\textbf{(I) Case: $v_{i_0}(t)\in[\Delta_2,\Delta_2+\varepsilon)$ and
there exists a directed path from $a_{i_p}$ to $a_{i_0}$.}

Let $a_{i_1},\dots,a_{i_p}$ denote those individuals whose shortest directed path
lengths to $a_{i_0}$ in $\mathbb{G}_\infty$ are $1,\dots,p$, respectively.
By \textbf{Theorem}~\ref{thm:convex_nonexpanding}, we have
$\Delta(t+\delta t)\subset\Delta(t)$ for all $\delta t>0$. Hence, for every
$t>t_k$ there exists some $a_i$ such that
$v_i(t)\in[\Delta_2,\Delta_2+\varepsilon)$.
Choose
\[
t_{k_0}=t_k+p(\tau_B+\tau_{min})>t_k
\]
so that
\[
v_{i_0}(t_{k_0})\in[\Delta_2,\Delta_2+\varepsilon),
\]
where $\tau_{min}>0$ and $\tau_B>0$ are, respectively, the lower bound on the
lengths of the time-invariant intervals of $\mathbb{G}(t)$ and the upper bound
on the maximal relaxation intervals for edges with infinite edge-length
measure.

By \textbf{Lemma}~\ref{lem:backwards}, for all
\[
t\in[t_{k_0}-(\tau_B+\tau_{min}),\,t_{k_0}]
\]
we have
\[
v_{i_0}(t)>\Delta_2-\varepsilon_0(\varepsilon),
\]
where $\varepsilon_0(\varepsilon)$ is an infinitesimal equivalent to
$\varepsilon$.
Then, by \textbf{Lemma}~\ref{lem:edge_prop}, there exists a time
\[
t_{k_1}\in[t_{k_0}-(\tau_B+\tau_{min}),\,t_{k_0}]
\]
such that
\[
v_{i_1}(t_{k_1})
>\Delta_2-\varepsilon'(\varepsilon_0(\varepsilon),\varepsilon)
= \Delta_2-\varepsilon_1(\varepsilon),
\]
where $\varepsilon_1(\varepsilon)$ is also an infinitesimal equivalent to
$\varepsilon$.

Repeating this procedure, we obtain a sequence of times
\[
\begin{array}{c}
   t_{k_0}=t_k+p(\tau_B+\tau_{min}),\quad
   t_{k_1}\in[t_{k_0}-(\tau_B+\tau_{min}),\,t_{k_0}],\quad
   \dots,\quad
   t_{k_p}\in[t_{k_{p-1}}-(\tau_B+\tau_{min}),\,t_{k_{p-1}}],
\end{array}
\]
such that
\[
\begin{array}{c}
   v_{i_0}(t_{k_0})>\Delta_2-\varepsilon_0(\varepsilon),\quad
   v_{i_1}(t_{k_1})>\Delta_2-\varepsilon_1(\varepsilon),\quad
   \dots,\quad
   v_{i_p}(t_{k_p})>\Delta_2-\varepsilon_p(\varepsilon),
\end{array}
\]
where $\varepsilon_0(\varepsilon),\varepsilon_1(\varepsilon),\dots,
\varepsilon_p(\varepsilon)$ are all infinitesimals equivalent to $\varepsilon$.

By \textbf{Lemma}~\ref{lem:backwards}, for all
\[
t\in[t_{k_{p-1}}-(\tau_B+\tau_{min}),\,t_{k_0}],
\ t\in[t_{k_{p-1}}-(\tau_B+\tau_{min}),\,t_{k_1}],
\ \dots,\
t\in[t_{k_{p-1}}-(\tau_B+\tau_{min}),\,t_{k_{p-1}}],
\]
we have
\[
\begin{array}{c}
   v_{i_0}(t)>\Delta_2-\varepsilon'_0(\varepsilon),\quad
   v_{i_1}(t)>\Delta_2-\varepsilon'_1(\varepsilon),\quad
   \dots,\quad
   v_{i_p}(t)>\Delta_2-\varepsilon'_p(\varepsilon),
\end{array}
\]
for certain infinitesimals $\varepsilon'_s(\varepsilon)$ equivalent to
$\varepsilon$.

Define
\[
\tilde\varepsilon(\varepsilon)
=\max\bigl\{\varepsilon'_0(\varepsilon),\varepsilon'_1(\varepsilon),
\dots,\varepsilon'_p(\varepsilon)\bigr\}.
\]
Then, at the common time
\[
t = t_{k_{p-1}} - (\tau_B+\tau_{min}),
\]
we obtain
\begin{equation}\label{th3-1}
v_{i_0}(t),\,v_{i_1}(t),\,\dots,\,v_{i_p}(t)
> \Delta_2 - \tilde\varepsilon(\varepsilon).
\end{equation}

\medskip
\noindent\textbf{(II) Case: $\mathbb{G}_\infty$ is strongly connected.}

Assume that the shortest directed path length from any vertex in $\mathbb{A}$
to $a_{i_0}$ does not exceed $p$. Since $\mathbb{G}_\infty$ is strongly
connected, then  every
$a_{i_s}\in\mathbb{A}$, $a_{i_s}\neq a_{i_0}$, has a directed path to
$a_{i_0}$ in $\mathbb{G}_\infty$. In this case, \eqref{th3-1} holds for the
states of all individuals at the common time
\[
t' = t_{k_{p-1}} - (\tau_B+\tau_{min}),
\]
so that
\[
\underline{\Delta}(t') > \Delta_2 - \tilde\varepsilon(\varepsilon).
\]
By \textbf{Theorem}~\ref{thm:convex_nonexpanding}, we have
$\Delta(t+\delta t)\subset\Delta(t)$ for all $\delta t>0$, and thus for all
$t\ge t'$,
\[
\underline{\Delta}(t) > \Delta_2 - \tilde\varepsilon(\varepsilon),
\qquad
\bigl|\underline{\Delta}(t)-\Delta_2\bigr|
=\Delta_2-\underline{\Delta}(t)
< \tilde\varepsilon(\varepsilon).
\]
Since $\tilde\varepsilon(\varepsilon)>0$ is an infinitesimal depending on
$\varepsilon$, letting $t\to\infty$ yields
\[
\underline{\Delta}(t)\to\Delta_1=\Delta_2.
\]
Hence the minimal convex cover $\Delta(t)$ collapses to a single point, and
the system reaches consensus.

\medskip
\noindent\textbf{(III) Case: $\mathbb{G}_\infty$ is not strongly connected, but
has a unique independent strongly connected component.}

Let $\mathbb{G}^1_\infty
=\langle\mathbb{A}_1,{\cal E}^1_\infty\rangle$ be the unique independent
strongly connected component of $\mathbb{G}_\infty$, and let $\mathbb{A}_1$ be
its corresponding independent basic set.
In $\mathbb{G}_\infty$ there is no
directed path from any vertex in $\mathbb{A}- \mathbb{A}_1$ to any
vertex in $\mathbb{A}_1$.
There exists $T>0$ such that, for
all $t>T$, there is no directed path in $\mathbb{G}(t)$ from any vertex in
$\mathbb{A} - \mathbb{A}_1$ to any vertex in $\mathbb{A}_1$.
Thus, for $t>T$, the dynamics
\begin{equation}\label{th3-2}
\dot v_j
= \sum_{a_s\in N(t,a_j)} -g_{js}\bigl(\|v_j-x_s\|\bigr)\,\vec r_{js},
\qquad a_j\in\mathbb{A}_1,
\end{equation}
constitutes an independent subsystem of \eqref{CS1}.
Its interaction network
\[
\mathbb{G}^1(t)
= \bigl\langle \mathbb{A}_1,\,
  (\mathbb{A}_1\times\mathbb{A}_1)\cap{\cal E}(t)
  \bigr\rangle
\]
has $\mathbb{G}^1_\infty=\langle\mathbb{A}_1,{\cal E}^1_\infty\rangle$ as its
$\infty$Cadjoint graph, which is strongly connected.

By case \textbf{(II)}, the states of all individuals in the subsystem
\eqref{th3-2} converge to some $\eta\in[\Delta_1,\Delta_2]$.
Assume, for contradiction, that $\eta\neq\Delta_2$.
Then for any $\varepsilon>0$ there exists $t_k$ such that, for all $t>t_k$,
\begin{equation}\label{th3-3}
\begin{array}{c}
   v_j(t)\in(\eta-\varepsilon,\ \eta+\varepsilon),
   \qquad a_j\in\mathbb{A}_1.
\end{array}
\end{equation}

At the same time, for
\[
t' = t_k + p(\tau_B+\tau_{min}),
\]
there exists $a_{i_0}$ such that
\[
v_{i_0}(t')\in[\Delta_2,\Delta_2+\varepsilon),
\]
where $p$ is an upper bound on the shortest directed path length between
vertices in $\mathbb{G}_\infty$.
Clearly, for sufficiently small $\varepsilon$, such an $a_{i_0}$ cannot belong
to $\mathbb{A}_1$.

Since $\mathbb{G}_\infty$ has a unique independent strongly connected
component $\mathbb{G}^1_\infty=\langle\mathbb{A}_1,{\cal E}^1_\infty\rangle$,
there
is a directed path from some $a_j\in\mathbb{A}_1$ to $a_{i_0}$.
Thus we may write $a_j=a_{i_s}$ for some $1\le s\le p$.
By \eqref{th3-1}, there exists $t_r>t_k$ (with
$t_r<t_k+p(\tau_B+\tau_{min})$) such that
\begin{equation}\label{th3-4}
v_j(t_r)=v_{i_s}(t_r) > \Delta_2 - \tilde\varepsilon(\varepsilon),
\end{equation}
where $\varepsilon$ and $\tilde\varepsilon(\varepsilon)$ are infinitesimals.

However, \eqref{th3-3} and \eqref{th3-4} are incompatible when
$\varepsilon$ is sufficiently small, which contradicts the assumption
$\eta\neq\Delta_2$.
A similar argument rules out the possibility $\eta\neq\Delta_1$.
Therefore $\eta=\Delta_1=\Delta_2$, and the system achieves state consensus.

\hfill$\Box$

%


%

\begin{thm}[Necessary condition]\label{thm:necessary}
Let the time-varying network $\mathbb{G}(t)$ be given. A necessary condition for the
velocity coordination system~\eqref{CS1} to achieve state consensus under arbitrary initial
conditions is that the $\infty$-adjoint graph $\mathbb{G}_\infty$ of $\mathbb{G}(t)$ contains exactly
one independent strongly connected component.
\end{thm}

\textbf{Proof.}

If $\mathbb{G}_\infty$ contains two independent strongly connected components, then these two components form two independent strongly connected subgraphs, corresponding to two independent subsystems.
By Theorem~\ref{thm:sufficient}, each subsystem will converge to its own consensus state.
For the overall system to converge to the \emph{same} limit under all initial conditions, the only possibility is that both subsystems converge to the same constant value for every initial condition.
This implies that the system must always converge to a fixed constant, completing the proof.

\hfill$\Box$

Combining Theorems~\ref{thm:sufficient} and \ref{thm:necessary}, we obtain the main structural
result of the paper.

\begin{thm}[Necessary and sufficient condition for consensus]\label{thm:nec_suff}
Assume that the time-varying communication network $\mathbb{G}(t)$ is a stationary time-varying
network. Then all individuals in the velocity coordination system~\eqref{CS1} reach state consensus
if and only if the $\infty$-adjoint graph $\mathbb{G}_\infty$ of $\mathbb{G}(t)$ contains a unique
independent strongly connected component.
\end{thm}

This theorem provides the topological condition under which large coordinated groups exist:
the presence of a single dominant SCC in the persistent interaction backbone. Biologically, this
dominant SCC is the subgroup that sets the velocity inherited by the rest of the group and by
any trailing subgroups that eventually merge into it.

%

\end{document}